\documentclass[prb,aps,twocolumn,showpacs]{revtex4}
\usepackage[dvips]{graphicx}
\usepackage{psfrag}

\begin{document}

\title{Functional renormalization group for nonequilibrium
  quantum many-body problems}

\author{R.\ Gezzi}
\affiliation{Institut f\"ur Theoretische Physik, Universit\"at G\"ottingen,
  Friedrich-Hund-Platz~1, D-37077 G\"ottingen, Germany}
\author{Th.\ Pruschke}
\affiliation{Institut f\"ur Theoretische Physik, Universit\"at G\"ottingen,
  Friedrich-Hund-Platz~1, D-37077 G\"ottingen, Germany}
\author{V.\ Meden}
\affiliation{Institut f\"ur Theoretische Physik, Universit\"at G\"ottingen,
  Friedrich-Hund-Platz~1, D-37077 G\"ottingen, Germany}

\date{\today}

\begin{abstract}
We extend the concept of the functional renormalization 
for quantum many-body problems to 
nonequilibrium situations. Using a suitable generating functional 
based on the Keldysh approach, we derive a system of coupled 
differential equations for the $m$-particle vertex
functions. The approach is completely general and  allows
calculations for both stationary and time-dependent situations.
As a specific example we study the stationary state transport
through a quantum dot with local Coulomb correlations at 
finite bias voltage employing two different
truncation schemes for the infinite hierarchy of 
equations arising in the functional renormalization group scheme.
\end{abstract}
\maketitle

\section{Introduction}
\label{sec:intro}
The reliable calculation of physical properties of interacting quantum
mechanical systems presents a formidable task. Typically, one has to 
cope with the interplay of different energy-scales possibly covering 
several orders of magnitude even for simple situations. Approximate 
tools like perturbation theory, but even numerically exact techniques 
can usually handle only a restricted window of energy scales and are 
furthermore limited in their applicability by the approximations 
involved or the computational resources available. In addition, due to 
the divergence of certain classes of Feynman diagrams, some of the 
interesting many-particle problems cannot be tackled by straight 
forward perturbation theory. 

The situation becomes even more involved if one is interested in properties
off equilibrium, in particular time-dependent situations. A standard approach
for such cases is based on the Keldysh formalism\cite{Keldysh} 
for the time evolution of Green functions, resulting in a 
matrix structure of propagators and self-energies.
This structure is a direct consequence of the fact
that in nonequilibrium we have to calculate averages of operators
taken not with respect to the ground state but with respect to
an arbitrary state. 
Therefore, the Gell-Mann and Low theorem\cite{fetterwalecka} is not
valid any more. 
Other approaches attempt to treat the time
evolution of the nonequilibrium system numerically, for example using
the density matrix renormalization
group\cite{schollw:rmp}, the numerical renormalization group 
(NRG)\cite{costi:97,anders:06}
or the Hamiltonian based flow-equation method.\cite{lobaskin:05}

The Keldysh technique shows a big flexibility and one therefore can 
find a wide range of applications such as transport through 
atomic, molecular and nano devices
under various conditions,\cite{Meir:1992}
systems of atoms
interacting with a radiation field in contact with a 
bath,\cite{koremann:66,langreth:76} or
electron-electron interaction in a weakly ionized
plasma.\cite{altshuler:78}

One powerful concept to study interacting many-particle systems 
is the rather general idea of the renormalization
group\cite{wilson:75} (RG), 
which has also been applied to time-dependent and stationary 
nonequilibrium situations 
recently.\cite{Schoeller:1999,Schoeller:2000,Keil:2001,Jens1,rosch:05}
In the RG approach one usually starts from high
energy scales, leaving out possible infrared divergences and works 
ones way down to the desired low-energy region in a systematic way. 
However, the precise definition of ``systematic way'' does in general depend 
on the problem studied.

In order to resolve this ambiguity for interacting quantum
mechanical many-particle systems in equilibrium, 
two different schemes attempting a unique, problem independent 
prescription have emerged during the past decade. One is Wegner's  
Hamiltonian based flow-equation technique,\cite{wegner,glazek} 
the second a field theoretical approach, which we want to focus on in
the following. This approach is based on a functional representation of the partition 
function of the system and has become known 
as functional renormalization group
(fRG).\cite{polchinski,wetterich,morris:94,salmhoferbuch}. 

A detailed 
description of the various possible implementations of the fRG and its
previous applications in equilibrium can be found
e.g.\ in Refs.~\onlinecite{Salmhofer:2001} and \onlinecite{hedden04}.
In the present work we extend the fRG 
to nonequilibrium by formulating the problem on the
real instead of the imaginary time axis using the functional integral
representation of the action on the Keldysh contour. 
Within a diagrammatic approach a similar set of fRG flow
equations has already been derived by Jakobs and Schoeller and applied to
study nonlinear transport through one-dimensional correlated electron systems.\cite{jakobs:03,jakobsneu}
We believe that this method will enable
us to treat a variety of nonequilibrium problems within a
scheme which is well established in equilibrium and in contrast to
other approaches is comparatively modest with respect to the computer 
resources required. Our framework for nonequilibrium will 
turn out to be sufficiently
general to allow for a treatment of systems disturbed by arbitrary
external fields (bias voltage, laser field, etc.), which can be constant or
time-dependent.

For classical many-body problems the fRG (also called nonperturbative
renormalization group) was used to study nonequilibrium 
transitions between stationary states.\cite{Canet}

As a simple but nontrivial application to test the potential and
weakness of our implementation of a nonequilibrium fRG we choose the single impurity 
Anderson model (SIAM).\cite{anderson:61} This model represents the
paradigm for correlation effects in condensed matter physics and is at
the heart of a large range of
experimental\cite{kastner92,kouwenhoven01,nygard00,madhavan98,li98,fiete03,pustilnik04}
and theoretical investigations.\cite{georges:96,maier:06}
It furthermore is the standard model for the  description of the 
transport properties of interacting single-level quantum dots.

The paper is organized as follows. In Sec.\ \ref{formalism} and the
appendices we extend the equilibrium fRG to treat general nonequilibrium
quantum many-body problems. As an example we study the 
finite bias, stationary transport through an Anderson 
impurity in Sec.\ \ref{appli}. 
Within the nonequilibrium extension we obtain a system of coupled tensor
equations, which represent the flow of the different components (in the 
Keldysh space) of the self-energy and the vertex function. We
discuss two approximations derived from the  fRG scheme that
have been successfully applied to the equilibrium 
situation.\cite{meden06:1,meden06:2} A summary and outlook in 
Sec.\ \ref{summary} concludes the paper.

\section{Functional RG in nonequilibrium}
\label{formalism}
We start from the standard definition of the two-time
Green function
\begin{equation}\label{eq:standard_GF} 
G (\xi',\xi)=-i \langle S^{-1} \psi_I(\xi')
  \psi_I^{\dag}(\xi) S \rangle\;\; ,
\end{equation} 
in the interaction picture, where  $\xi$  comprises a
set of single-particle quantum numbers and time $t$. 
The time evolution operators are
given as
\begin{eqnarray}\label{eq:time_evolve}
S & = & {\rm T} \exp\left\{-i\int\limits^{\infty}_{-\infty}V_I(t)dt\right\}\\[5mm]
\label{eq:anti_time_evolve}
S^{-1} & =& \tilde{\rm T} \exp\left\{i\int\limits^{\infty}_{-\infty}V_I(t)dt\right\}\;\;,
\end{eqnarray} 
with ${\rm T}$ the usual time ordering and $\tilde{\rm T}$
the anti time-ordering operator. The interaction term $V_I(t)$
is arbitrary, including possible explicit time dependence. 

\begin{figure}[tb]
  \begin{center}
\psfrag{Ck+}[cb][cb]{$C_{K_-}$}
\psfrag{Ck-}[ct][ct]{$C_{K_+}$}
\psfrag{t}[ct][ct]{$t$}
\psfrag{minf}[rc][rc]{$-\infty$}
\psfrag{pinf}[lc][lc]{$\infty$}
    \includegraphics[width=0.4\textwidth,clip]{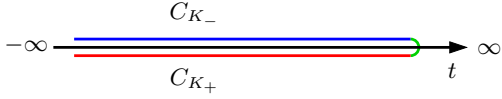}
  \end{center}
  \caption[]{(color online) Keldysh contour}
  \label{fig:keldysh_contour}
\end{figure}

In a nonequilibrium situation, the propagation of the system from
$-\infty\to\infty$ is not any more equivalent to the propagation
from $\infty\to-\infty$, i.e.\ one has to distinguish whether the time
arguments in Eq.\ (\ref{eq:standard_GF}) belong to the former or
latter.\cite{LandauX,Rammer,Haug} This scheme is usually depicted by
the Keldysh double-time contour shown in Fig.\ \ref{fig:keldysh_contour}, where
$C_{K_-}$ represents the propagation $-\infty\to \infty$ (upper branch
of the Keldysh contour) and $C_{K_+}$
the propagation $\infty\to-\infty$ (lower branch
of the Keldysh contour).
Consequently, one has to introduce four
distinct propagators, namely the time-ordered Green function
\begin{eqnarray}
\label{eq:time_o_GF} 
\!\!\!\!\!\!\!\!\!\!\! && G^{--}(\xi',\xi) = -i\langle {\rm T} \psi(\xi')\psi^{\dag}(\xi) \rangle \nonumber \\*
\!\!\!\!\!\!\!\!\!\!\! && = -i\theta(t'-t)\langle\psi(\xi')\psi^{\dag}(\xi)\rangle-\zeta i\theta(t-t')\langle
\psi^{\dag}(\xi)\psi(\xi') \rangle ,
\end{eqnarray}
where $t,t'\in C_{K_-}$, the anti time-ordered Green function 
\begin{eqnarray}
\label{eq:anti_time_o_GF}
\!\!\!\!\!\!\!\!\!\!\! && G^{++}(\xi',\xi) =  -i\langle \tilde{\rm T} \psi(\xi') \psi^{\dag}(\xi) \rangle
\nonumber \\
\!\!\!\!\!\!\!\!\!\!\! && = -i\theta(t-t')\langle
\psi(\xi')\psi^{\dag}(\xi)\rangle-\zeta i\theta(t'-t)\langle
\psi^{\dag}(\xi)\psi(\xi') \rangle ,
\end{eqnarray}
with $t,t' \in C_{K_+}$ and 
\begin{eqnarray}\label{eq:Gmp}
G^{+-}(\xi',\xi) &=& -i\langle
\psi(\xi')\psi^{\dag}(\xi)\rangle\;,t'\in C_{K_+}, t \in C_{K_-}\\
\label{eq:Gpm}
G^{-+}(\xi',\xi) &=& -\zeta i\langle \psi^{\dag}(\xi)\psi(\xi')\rangle\;,t'\in
C_{K_-},t\in C_{K_+}
\end{eqnarray}
where $\zeta=+1$ for bosons and $\zeta=-1$ for fermions.
Concerning the Keldysh indices we here follow the notation 
of Ref.\ \onlinecite{LandauX}. 
$G^{--}(\xi',\xi)$ and $G^{++}(\xi',\xi)$ take into account the
excitation spectrum while $G^{+-}(\xi',\xi)$ and $G^{-+}(\xi',\xi)$ describe the thermodynamic state of
the system.
Only three of the Green functions are independent and one commonly 
introduces the linear combinations
\begin{eqnarray}\label{eq:Gret}
G^{\rm R}(\xi',\xi) &=&
\theta(t'-t)\left[G^{+-}(\xi',\xi)+\zeta G^{-+}(\xi',\xi)\right]\\
\label{eq:Gadv}
G^{\rm A}(\xi',\xi) &=&
\theta(t-t')\left[G^{-+}(\xi',\xi)+\zeta G^{+-}(\xi',\xi)\right]\\
\label{eq:Gkeldysh}
G^{\rm K}(\xi',\xi) &=& G^{-+}(\xi',\xi)+G^{+-}(\xi',\xi)\;\;,
\end{eqnarray}
named retarded, advanced and Keldysh component, respectively.

To derive a nonequilibrium fRG scheme for an interacting many-body
problem along the lines of 
Refs.~\onlinecite{Salmhofer:2001} and \onlinecite{hedden04} one
needs a formulation that allows
to express the $m$-particle vertices as functional derivatives of a generating
functional. Here we use the approach by Kameneev.\cite{kameneev}
To set up a functional integral representation of the generating
functional we define the matrix  
$$
\hat{G}=\left(\begin{array}{cc}G^{--} & G^{-+}\\
G^{+-} & G^{++}\end{array}\right)
$$
and the short hand notation 
\begin{eqnarray*}
\left(\bar\psi,\hat{O}\psi\right) = i\int\limits_{-\infty}^\infty
d\xi d\xi' 
\bar\psi(\xi)\hat{O}_{\xi,\xi'}\psi(\xi')\;\;,
\end{eqnarray*}
where 
$$
\psi =
\left( \begin{array}{c}
\psi_{-}\\
\psi_{+} 
\end{array}\right)
$$
is a spinor of fields (Grassmann for fermions or complex for
bosons) with $\psi_-$ having a time from the upper branch of the
Keldysh contour and  $\psi_+$ a time from the lower.
Later it will also prove useful 
to Fourier transform from time $t$ to frequency $\omega$. 
One then has to replace $t$ in $\xi$ by $\omega$. The integrals over 
$\xi$ and $\xi'$ stand for summations over the quantum numbers and  
integrations over time or frequency.
The following steps can be performed with $\xi$ either containing time
or frequency.   
The generalization of the functional integral representation 
of the partition function to nonequilibrium is\cite{kameneev}
\begin{widetext}
\begin{eqnarray}
\label{eq:partitionfunction}
\Xi = \frac{1}{\Xi_0} \int
{\mathcal D} \bar \psi \psi \exp{\left\{ \left(\bar \psi,
    \left[\hat{G}_0\right]^{-1}\psi \right) - iS_{\rm int}\left(\{ \bar \psi
      \}, \{\psi\}\right) \right\} } \;, \;\; \Xi_0= \int{\mathcal D} \bar \psi \psi
\exp{\left\{\left(\bar\psi,\left[\hat{G}_0\right]^{-1}\psi\right)\right\}}\;\;.
\end{eqnarray}
The matrix $\hat{G}_0$ denotes the propagator of the 
noninteracting part of the many-body problem and $S_{\rm int}$ 
is the (arbitrary) interaction term.

To construct a generating functional for $m$-particle Green
functions, one introduces external source fields $\eta$ and $\bar\eta$
\begin{equation}
\label{eq:genfunct}
{\mathcal W } \left(\{\bar \eta \}, \{ \eta\} \right) = 
\frac{1}{\Xi_0} 
\int {\mathcal D} \bar \psi \psi  \exp \bigg\{ 
 \left( \bar \psi, \left[\hat{G}_{0}\right]^{-1}
   \psi\right) - iS_{\rm int}(\{\bar \psi \}, \{ \psi \})
 - \left( \bar \psi, \eta\right) - \left(\bar \eta,  \psi  
   \right) \bigg\} \;.
\end{equation}
The (connected) $m$-particle Green function $G_m^{(c)}$ can then be
obtained by taking functional derivatives 
\begin{equation}
\label{eq:gmdef}
G_m^{(c)}\left(\left\{\xi_j'\right\};\left\{\xi_j\right\}\right) = 
\left.(\zeta i)^m \frac{\delta^m}{\delta \bar \eta_{\xi_1'} 
\ldots \delta \bar \eta_{\xi_m'}} 
\frac{\delta^m}{\delta  \eta_{\xi_m} \ldots  \delta  \eta_{\xi_1}} {\mathcal
  W}^{(c)}\left(\{\bar \eta \}, \{ \eta\} \right) \right|_{\eta
=\bar\eta=0}
\end{equation}
with 
\begin{eqnarray}
\label{eq:wcdef}
{\mathcal W}^c\left(\{\bar \eta \}, \{ \eta\} \right) = 
\ln{ \left[ {\mathcal W}
\left(\{\bar \eta \}, \{ \eta\} \right) \right]} \; .
\end{eqnarray}
The derivatives in Eq.\ (\ref{eq:gmdef}) are taken with respect
to the spinors $\eta$ and $\bar \eta$, i.e.\ the resulting Green function 
$G_m^{(c)}$ is a tensor of rank $2m$ in the Keldysh 
indices. In the following we adopt
the notation that quantities with explicit index $m$ are tensors of
rank $2m$; without index $m$ we denote the propagator and
self-energy, both carrying a hat to point out their matrix structure.

Introducing the fields
\begin{eqnarray}\label{eq:new_fields}
\phi_\xi =  i\frac{\delta}{\delta \bar \eta_\xi} {\mathcal
  W}^c\left(\{\bar \eta \}, \{ \eta\} \right) \; , \;\;\;\; 
\bar\phi_\xi =  \zeta i \frac{\delta}{\delta \eta_\xi} {\mathcal
  W}^c\left(\{\bar \eta \}, \{ \eta\} \right)
\end{eqnarray}
we can perform a Legendre transformation
\begin{eqnarray}
\label{eq:gammadef}
\Gamma \left(\{\bar \phi \}, \{ \phi \} \right) = -  {\mathcal
  W}^c\left(\{\bar \eta \}, \{ \eta\} \right) -\left(\bar \phi, \eta
  \right)
-\left(\bar \eta, \phi \right)
 + \left(\bar \phi, \left[\hat{G}_{0}\right]^{-1}\phi\right)  \; , 
\end{eqnarray} 
to the generating functional of the one-particle irreducible vertex 
functions $\gamma_m$ (tensors of rank $2m$ in the Keldysh indices)
\begin{equation}
\label{eq:vertexfundef}
\gamma_m\big(\left\{\xi_j'\right\};\left\{\xi_j\right\}\big) =
\left. i^m \frac{\delta^m}{\delta \bar \phi_{\xi_1'} \ldots 
\delta \bar \phi_{\xi_m'}}
\frac{\delta^m}{\delta  \phi_{\xi_m} \ldots \delta  \phi_{\xi_1}} \Gamma
\left(\{\bar \phi\}, \{ \phi\} \right) \right|_{\phi=\bar\phi=0}\;.
\end{equation}
\end{widetext}
In contrast to the usual
definition\cite{negele} of $\Gamma$ for convenience (see Appendix A)
we added a term $\left(\bar \phi, \left[\hat{G}_{0}\right]^{-1}
  \phi\right)$ in Eq.\ (\ref{eq:gammadef}). The relations between the
$G_m^{(c)}$ and $\gamma_m$ in imaginary time can be found in text
books\cite{negele} and can straightforwardly be extended to real times
on the Keldysh contour.
For example, for the one-particle Green function one obtains
\begin{eqnarray*}
G_1(\xi';\xi) & = & G_1^c(\xi';\xi)\\ 
& = & i \frac{\delta}{\delta \bar \eta_{\xi_1'}} 
\frac{\delta}{  \delta  \eta_{\xi_1}} {\mathcal W}^c = -\zeta
\hat{G}_{\xi',\xi}\;\; , \\
& = &
  \left[  \gamma_1 - \zeta \left[\hat{G}_0\right]^{-1}
  \right]^{-1}_{\xi',\xi} 
\end{eqnarray*}
where
\begin{eqnarray*}
\hat{G}_{\xi',\xi}  =
\left(\left[\hat{G}_0\right]^{-1} - \hat{\Sigma}\right)^{-1}_{\xi',\xi} ,
\end{eqnarray*}
with the proper one-particle self-energy $\hat{\Sigma}$. This implies the relation $\hat{\Sigma} = \zeta \gamma_1$.
We have a matrix structure  not only with respect to
$\xi$ and $\xi'$ (as in equilibrium), but also with respect to the 
Keldysh indices. How does this additional structure manifests 
itself in the Green functions (\ref{eq:gmdef})?
For  $m=1$ we have two derivatives with respect to spinor
fields, i.e.\ the structure of a tensor product, which can be made
explicit by using a tensor product notation 
$$
G_1^{c}\left( \xi',\xi\right) =
(\zeta i)\left.\frac{\delta}{\delta \bar \eta_{\xi'}} \otimes\frac{\delta}
{\delta\eta_{\xi}}{\mathcal W}^c\right|_{\eta=\bar\eta=0}\;\;,
$$
leading to the matrix
\begin{eqnarray}
\label{eq:gm_one_p}
G_1^{c}\left( \xi',\xi\right) &=&
(\zeta i)\left(\begin{array}{cc}\frac{\delta^2 {\mathcal W}^c}
{\delta\bar\eta_{-}(\xi') \delta \eta_{-}(\xi)} &\frac{\delta^2
  {\mathcal W}^c }{\delta \bar \eta_{-}(\xi')\delta 
\eta_{+}(\xi)}  \\[2mm]  \frac{\delta^2 {\mathcal W}^c }
{\delta \bar \eta_{+}(\xi')\delta  \eta_{-}(\xi)} & \frac{\delta^2
  {\mathcal W}^c }{\delta \bar \eta_{+}(\xi')\delta  \eta_{+}(\xi)}
\end{array}\right)\nonumber\\[2mm] 
&=&
\left(\begin{array}{cc} G^{--}(\xi',\xi) &  G^{-+}(\xi',\xi) \\[2mm] 
 G^{+-}(\xi',\xi) &  G^{++}(\xi',\xi)
\end{array}\right)\;\;.
\end{eqnarray}

Up to now we extended standard text book 
manipulations\cite{negele} from imaginary time to 
the real-time Keldysh contour. We emphasize that no translational 
invariance in time is assumed. To derive the fRG flow
equations in nonequilibrium we can 
follow the steps of Ref.~\onlinecite{hedden04}. In 
Eqs.\ (\ref{eq:partitionfunction}) and (\ref{eq:genfunct}) we replace the
noninteracting propagator by a propagator $\hat{G}_{0}^\Lambda$
depending on a parameter $\Lambda \in [\Lambda_0,0]$ and require
\begin{eqnarray}
\label{eq:demands}
\hat{G}_{0}^{\Lambda_0} =  0  & , & \hat{G}_{0}^{\Lambda=0} = \hat{G}_{0} \; ,
\end{eqnarray} 
i.e.\ at the starting point $\Lambda = \Lambda_0$ no degrees of
freedom are ``turned on'' while at $\Lambda=0$ the $\Lambda$ free
problem is recovered. In models with infrared divergences $\Lambda$
can be used to regularize the problem. In equilibrium this is often
be achieved by implementing $\Lambda$ as an infrared cutoff in momentum
or energy. One of the advantages of the fRG approach over other RG
schemes is that one is not restricted to these choices and other ways
of introducing the parameter $\Lambda$ have turned out to be 
useful for equilibrium problems.\cite{Hon,lecturenotes} All that is 
required to derive the fundamental flow equations (\ref{eq:gamma1f2}) 
and (\ref{eqq1}) (see below) are the conditions Eq.\
(\ref{eq:demands}). In our application of the nonequilibrium fRG to
the steady state transport through an
interacting quantum dot it is natural to implement $\Lambda$ as an energy
cutoff. However, such a choice must not be the natural one in cases
where one is interested in studying transient
behavior. In this situation  the propagator and the vertex functions 
in general depend on several spatial and time variables and there is
no obvious momentum or energy cutoff scheme. Within the fRG several
ways of introducing $\Lambda$ can be worked out, compared and the one
best suited for the problem under investigation can be identified.  

Through $\hat{G}_{0}^\Lambda$ the quantities defined in 
Eqs.\ (\ref{eq:partitionfunction}) to
(\ref{eq:vertexfundef}) acquire a $\Lambda$-dependence. 
Taking the derivative with respect to $\Lambda$ one now derives 
a functional differential equation for $\Gamma^{\Lambda}$. From this, 
by expanding in powers of the 
external sources, an infinite hierarchy of coupled differential equations for the 
$\gamma_m^\Lambda$ is obtained. Although the steps in the derivation
are formally equivalent to
Ref.~\onlinecite{hedden04}, because of the real-time formulation
additional factors $i$ and signs appear in several places. We thus
believe that it is helpful to present the details of the derivation
for the present approach, which is done in Appendix A.

\begin{figure}[tb]
\begin{center}
\psfrag{k}[br][br]{$\xi$}
\psfrag{k'}[br][br]{$\xi'$}
\includegraphics[width=0.2\textwidth]{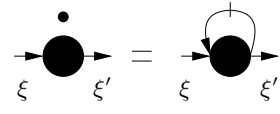}
\end{center}
\caption{\label{fig:gamma1} Diagrammatic form of the flow equation for
 $\gamma_1^{\Lambda}$. The slashed line
 stands for the single scale propagator $\hat{\mathcal S}^{\Lambda}$.}
\end{figure}

In particular, for the flow of the self-energy one finds the expression
\begin{eqnarray}
\frac{d}{d \Lambda} \gamma_1^{\Lambda}(\xi';\xi) &=& 
\zeta \frac{d}{d \Lambda} \hat{\Sigma}^{\Lambda}_{\xi',\xi}\nonumber\\
\label{eq:gamma1fl}
&=& \mbox{ Tr}\, \left[\hat{\mathcal S}^{\Lambda} 
 \gamma_2^{\Lambda}(\xi', \cdot; 
\xi, \cdot) \right] \; , 
\end{eqnarray}  
which can be visualized by the diagram in Fig.~\ref{fig:gamma1}.
The trace in Eq.~(\ref{eq:gamma1fl}) is meant to run over all 
quantum numbers and time respectively frequency.
In Eq.~(\ref{eq:gamma1fl}) appears the so-called single scale propagator
(the slashed line in Fig.~\ref{fig:gamma1})
\begin{eqnarray}
\label{eq:Sdef}
\hat{\mathcal S}^{\Lambda}  & = &
\hat{G}^{\Lambda} \hat{\mathcal Q}^{\Lambda}   
\hat{G}^{\Lambda}  \;,\\
\label{eq:Qlambdadef}
\hat{\mathcal Q}^{\Lambda} &=& \frac{d}{d \Lambda} \left[ \hat{G}_{0}^\Lambda\right]^{-1} \; .
\end{eqnarray}  
and the quantity $\gamma_2^{\Lambda}(\xi', \cdot; 
\xi, \cdot)$ denotes the matrix obtained by keeping the indices $\xi$ and $\xi'$ fixed.

\begin{figure}[tb]
\begin{center}
\psfrag{k}[br][br]{$\xi$}
\psfrag{k'}[br][br]{$\xi'$}
\includegraphics[width=0.45\textwidth]{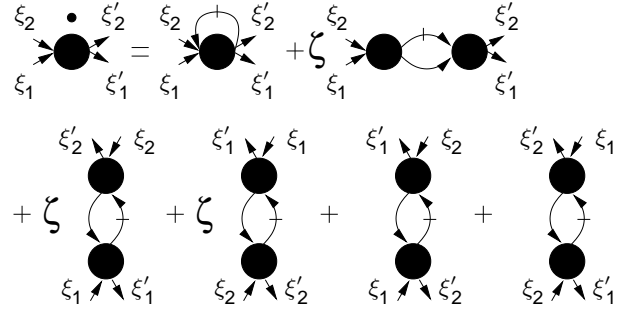}
\end{center}
\caption{\label{fig:gamma2} Diagrammatic form of the flow equation for
 $\gamma_2^{\Lambda}$. The slashed line
 stands for the single scale propagator $\hat{\mathcal S}^{\Lambda}$,
the unslashed line for $\hat{G}^{\Lambda}$.}
\end{figure}

We thus arrive at an expression that is formally identical to
Eq.~(19) in Ref.~\onlinecite{hedden04}. The difference appears in the
matrix structure, which now also contains the index components for the
branches of the Keldysh contour. To make this explicit, we write out
Eq.~(\ref{eq:gamma1fl}) with respect to the Keldysh structure
\begin{equation}\label{eq:gamma1f2}
\zeta\frac{d}{d\Lambda}\Sigma^{\alpha\beta,\Lambda}_{\xi',\xi}
=
{\rm Tr}\sum\limits_{\mu\nu}{\mathcal S}^{\mu\nu,\Lambda}
\gamma_{2}^{\alpha\nu\beta\mu,\Lambda}(\xi',\cdot;\xi,\cdot)\;,
\end{equation}
where $\alpha,\beta,\mu$ and $\nu$ denote Keldysh indices and take the
values $\mp$.  
Apparently, the derivative of $ \gamma_1^{\Lambda}$ is determined by $
\gamma_1^{\Lambda}$  and the
two-particle vertex $ \gamma_2^{\Lambda}$. Thus an equation for $
\gamma_2^{\Lambda}$ is required.
The structure of this equation turns out to be similar to 
Eq.~(21) of Ref.~\onlinecite{hedden04}. Here we only show the diagrams
representing it in Fig.~\ref{fig:gamma2}, the full expression is given
in Eq.\ (\ref{eqq1}).
The differential equation for $\gamma_2^\Lambda$ does not only contain 
$\gamma_1^{\Lambda}$ -- implicitly via the propagators -- and
$\gamma_2^{\Lambda}$, but also the three-particle vertex 
$\gamma_3^{\Lambda}$. The flow of the three-particle 
vertex depends on the four-particle vertex etc.
It is generically impossible to solve the full set of infinitely many
coupled differential equations. In applications one has to
truncate it, and this is usually done at order $m=2$, i.e.\ one replaces all
vertices with $m>2$ by their initial values, which for problems
with a two-particle interaction means $\gamma_m^\Lambda=\gamma_m^{\Lambda_0}=0$ for
$m>2$. Even within this truncated system the remaining set of differential
equations must typically further be approximated to allow for an
analytical or a  
numerical solution.\cite{Honerkamp:2003,meden06:1,meden06:2}
At the end of the fRG flow, that is for $\Lambda=0$, the
$\gamma_m^{\Lambda=0}=\gamma_m$ present approximations to the many-body
vertex functions from which observables can be computed (for examples
see Sec.\ \ref{appli}).

We again note that a similar set of flow equations 
was derived by Jakobs and Schoeller using a 
diagrammatic approach.\cite{jakobs:03,jakobsneu} 

\section{Application to quantum dots}
\label{appli}
The derivation of the flow equations in the previous section and the
appendices was completely 
general and they can be used to study time
dependent or stationary properties of systems at finite temperature $T$
or $T=0$. A simple but nonetheless nontrivial application
is nonequilibrium, stationary transport through an interacting
single-level quantum dot at
$T=0$. For a stationary situation, time translational invariance holds and one can 
Fourier transform to frequency space. Moreover, because the quantum dot
is a zero-dimensional structure, no additional degrees of freedom except spin
have to be taken into account, thus considerably reducing the complexity of the
problem. 

Nonequilibrium theory of single-level quantum dots described by the 
SIAM (see below) has been a major
subject of research over the past years. Several techniques have been
developed respectively applied, for example
statistical\cite{goglin:05,han:06,han:06_2}, 
perturbative,\cite{Hershfield:1991,hershfield:92,meir:92,Yeyati,wingreen:94,fujii:03,han:04,hamasaki:05,datta:06,Thygesen} RG
based\cite{Schoeller:2000,Jens1,rosch:05} and numerical
procedures.\cite{lobaskin:05,anders:06} In particular the RG based
calculations were up to now restricted to situations away
from the Kondo regime, either by looking at features in the
mixed-valence state \cite{Schoeller:2000} or in strong magnetic fields.\cite{Jens1,rosch:05} Despite
this tremendous ongoing effort, relatively little is known about
nonequilibrium properties, even in the stationary state, of the model. On the
other hand, the very detailed knowledge of the {\em equilibrium}
properties\cite{hewson:book} makes it possible to interpret certain
features or even dismiss approximations on the basis of this
fundamental understanding.
\subsection{Single impurity Anderson model}
The standard model to describe transport through
interacting quantum dots is the SIAM.\cite{anderson:61}
Experimentally,\cite{kastner92,kouwenhoven97} the dot region is attached to two external leads, and the
current is driven by an applied bias voltage $V_B$. Furthermore, the filling of the dot
can be controlled by a gate voltage $V_G$. To keep the notation simple we
use units with $e=\hbar=1$. The Hamiltonian reads
\begin{eqnarray}
H
& = &
\sum_{\vec{k}\sigma l} \varepsilon_{\vec{k}\sigma l}
c^\dag_{\vec{k}\sigma l} c^{\phantom{\dag}}_{\vec{k}\sigma l}
\nonumber\\
&& + \sum_\sigma V_G d^\dag_\sigma d^{\phantom{\dag}}_\sigma + 
U\left(n_\uparrow-\frac{1}{2}\right)\left(n_\downarrow-\frac{1}{2}\right)
\nonumber\\
\label{eq:siam} 
&& +\sum_{\vec{k}\sigma
  l}\left[V_{\vec{k}\sigma l}c^\dag_{\vec{k}\sigma
    l}d^{\phantom{\dag}}_\sigma+
\mbox{H.c.}\right] 
\end{eqnarray}
in standard second quantized 
notation ($n_\sigma =  d^\dag_\sigma d^{\phantom{\dag}}_\sigma$).
As usual, $\vec{k}$ is the wave vector of the band states in the leads and
$\sigma$ the spin. In addition, the index
$l=L,R$ distinguishes the left and right reservoirs which can
have different chemical potentials $\mu_l$ through an applied
bias voltage $V_B=\mu_L-\mu_R$. 
Since we do not include a magnetic field lifting the spin-degeneracy
of the spin up and down level, the one-particle Green
function and the self-energy are spin independent and we suppress the
spin index.  We assume the dispersions
$\varepsilon_{\vec{k}\sigma l}$ and hybridizations
$V_{\vec{k}\sigma l}$ between dot and the left and right
leads to be identical, and $V_{\vec{k}}\equiv V/\sqrt{2}$ to be
$\vec{k}$-independent. Equation (\ref{eq:siam}) is written such that
for $V_G=0$ we have particle-hole symmetry.

The interaction in the Hamiltonian is reduced to the dot site only. We
can thus reduce the problem to a zero-dimensional one by ``integrating
out'' the leads.  For the flow equations we need the propagator $\hat{G}_{d,0}$
at $U=0$. To this end we use the
Dyson equation to obtain\cite{Keldysh,LandauX,Haug}
\begin{equation}\label{eq:kop1} 
\hat{G}_{d,0}(\omega)=\left[ \left(\begin{array}{cc}
\omega-V_G & 0\\
0 & -\omega+V_G\end{array}\right)
-\hat{\Sigma}_{{\rm lead}} \right]^{-1}\;\;,
\end{equation}
where
\begin{eqnarray}
\hat{G}_{d,0} &=& 
\left(\begin{array}{cc} G^{--}_{d,0} & G^{-+}_{d,0}  \\[2mm]  
G^{+-}_{d,0} & G^{++}_{d,0} 
\end{array}\right)\;,\nonumber\\
\label{eq:hybridization}
\hat{\Sigma}_{\rm lead}
&=&
\frac{V^2}{2N}\sum_{\vec{k} l} 
\left(\begin{array}{cc}
G^{--}_{\vec{k} l} & G^{-+}_{\vec{k} l} \\[2mm]
G^{+-}_{\vec{k} l} & G^{++}_{\vec{k} l} 
\end{array}\right)\;\;.
\end{eqnarray}
To further evaluate Eq.\ (\ref{eq:hybridization}),
we have to insert the Green functions of the free electron gas, given by
\begin{eqnarray}
\label{eq:Gkpp} 
G^{--}_{\vec{k} l}(\omega)
&=&
\begin{array}[t]{l}\displaystyle
\frac{1}{\omega-\varepsilon_{\vec{k}}+\mu_l+i\delta} \\[5mm]
\displaystyle
+2i\pi f(\varepsilon_{\vec{k}})
\delta(\omega-\varepsilon_{\vec{k}}+\mu_l)\;,\end{array}\\
G^{++}_{\vec{k} l}(\omega)
&=&
-\left[G_{\vec{k} l}^{--}(\omega)\right]^{*}\;,\\
G_{\vec{k} l}^{-+}(\omega)
&=&
2i\pi f(\varepsilon_{\vec{k}})\delta(\omega-\varepsilon_{\vec{k}}+\mu_l)\;,\\
G_{\vec{k} l}^{+-}(\omega)
&=&
-2i\pi f(-\varepsilon_{\vec{k}})
\delta(\omega-\varepsilon_{\vec{k}}+\mu_l)\; ,
\end{eqnarray}
into Eq.~(\ref{eq:kop1}). This leads to
\begin{eqnarray}\label{eq:kulp1}G_{d,0}^{--}(\omega) &=&
\frac{\omega-V_G-i\Gamma\left[1 - 
  f_L(\omega)- f_R(\omega)\right]} {(\omega-V_G)^2+\Gamma^2}\;, \\
\label{eq:kulp2}G_{d,0}^{++}(\omega)&=&-[G_{d,0}^{--}(\omega)]^*\;,\\
\label{eq:pulp1}G_{d,0}^{-+}(\omega)&=&i\frac{\Gamma\left[
  f_L(\omega)+f_R(\omega)\right]}{(\omega-V_G)^2+\Gamma^2}\;,\\
\label{eq:tulp1}G_{d,0}^{+-}(\omega)&=&-i\frac{\Gamma\left[
  f_L(-\omega)+f_R(-\omega)\right]}{(\omega-V_G)^2+\Gamma^2}\;,
\end{eqnarray}
where $f_l(\pm\omega)=f\left(\pm(\omega-\mu_l)\right)$ 
are the Fermi functions of the leads and
$\Gamma=\pi |V|^2 N_F$, with $N_F$ the density of states at the Fermi
level of the semi-infinite leads, represents the tunnel barrier 
between the leads and
the impurity. 

Finally, the quantities we want to calculate are the current
$J$ through the dot and the differential conductance $G=dJ/dV_B$. For
the model (\ref{eq:siam}) the current is given
by\cite{Meir:1992,wingreen:94} 
\begin{widetext}
\begin{eqnarray}
  \label{eq:current}
  J = \frac{1}{2}(J_L+J_R)=
 \frac{i \Gamma}{2\pi}\int
  d\epsilon\left[f_L(\epsilon)-f_R(\epsilon)\right] \left[G^{+-}_d(\epsilon)-G^{-+}_d(\epsilon)\right]\;,
\end{eqnarray}
where we already used that the left and right couplings are identical,
i.e.\ $\Gamma_L=\Gamma_R=\Gamma/2$, and summed over both spin
directions. The interacting one-particle Green function of the dot is
denoted by $\hat G_d$ and $J_{L/R}$ are the currents across the left
and right dot-lead contacts respectively. Equation (\ref{eq:current})
is written in a somewhat unusual form, not employing the relation
$G^{+-}_d(\epsilon)-G^{-+}_d(\epsilon)=G^R_d(\epsilon)-G^A_d(\epsilon)=-2\pi i\rho_d(\epsilon)$, where $\rho_d$ denotes the dot's one-particle spectral function. The reason
for this will be explained below.

Another quantity of interest is $\Delta
J=J_L-J_R$.\cite{hershfield:92} 
Obviously, since no charge is produced on the quantum dot, 
$\Delta J=0$ in the exact solution. Using again the results of
Refs.~\onlinecite{Meir:1992} and \onlinecite{wingreen:94}, the
expression for $\Delta J$ becomes
\begin{eqnarray}
  \label{eq:DeltaJ2}
  \Delta J &=& 
-\frac{\Gamma}{\pi}\int\limits_{-\infty}^\infty d\omega
\frac{F(\omega)\left[\Im m\Sigma^{-+}(\omega)-\Im
  m\Sigma^{+-}(\omega)\right]-2\Im
m\Sigma^{-+}(\omega)}{\tilde\Delta(\omega)}\\[2mm]
\tilde\Delta(\omega) &=&
\left|\omega-V_G+i\Gamma\left[1-F(\omega)\right]-\Sigma^{--}(\omega)\right|^2+\left[\Gamma
  F(-\omega)+\Im m\Sigma^{+-}(\omega)\right]\left[\Gamma
  F(\omega)-\Im m\Sigma^{-+}(\omega)\right]\nonumber\\[2mm]
\label{eq:Fofw}
F(\omega) &=& f_L(\omega)+f_R(\omega)
\end{eqnarray}
\end{widetext}
where we used that $\Sigma^{-+}(\omega)$ and $\Sigma^{+-}(\omega)$ are
purely imaginary. Depending on the type of approximation used $\Delta
J=0$ might either hold for all parameters\cite{wingreen:94,Thygesen}
or not.\cite{hershfield:92}
We note that fulfilling $\Delta J=0$ is, however, not sufficient for an
approximation to provide reliable results. E.g.\ the self-consistent 
Hartree-Fock approximation fulfills $\Delta J=0$, but nonetheless
does not capture the correct physics
even in equilibrium.\cite{hewson:book}    

\subsection{Lowest order approximation}

Before studying the coupled system for $\hat\Sigma^\Lambda$ and $\gamma_2^\Lambda$ 
we begin with the simpler case where we replace $\gamma_2^\Lambda$ 
on the right hand side of Eq.\ (\ref{eq:gamma1f2}) by the
antisymmetrized bare interaction and consider only the flow of
$\hat\Sigma$. The only nonzero components 
of the bare two-particle vertex are those with all four Keldysh
indices being the same ($\alpha=\mp$)\cite{LandauX} 
\begin{eqnarray}
\label{eq:vertexinit}
\gamma^{\alpha\alpha\alpha\alpha,\Lambda_0}_{2}(\xi_1',\xi_2';\xi_1,\xi_2)
= \alpha iU
\left(\delta_{\sigma_1,\sigma_1'}\delta_{\sigma_2,\sigma_2'}-
\delta_{\sigma_1,\sigma_2'}\delta_{\sigma_2,\sigma_1'}
\right)  \; .
\end{eqnarray}
With this replacement Eq.\ (\ref{eq:gamma1f2}) reduces to
\begin{equation}
\label{eq:pot}
\frac{d}{d\Lambda}{\Sigma}^{\mp\mp,\Lambda} 
=\pm iU\int\frac{d\omega}{2\pi}{\mathcal S}^{\mp\mp,\Lambda}(\omega)\;\;.
\end{equation}
Within this approximation the self-energy is always time
or frequency independent, and no terms off-diagonal in the Keldysh contour
indices are generated.
It leads to at least qualitatively good
results in equilibrium,\cite{meden06:1} and has the additional
advantage that the flow equations can be solved analytically.

As the last step we specify how the parameter $\Lambda$
is introduced. Since we are interested in a stationary situation, i.e.\
the propagators only depend on the time difference $t-t'$, all
equations can be transformed into frequency space and one 
natural choice is a frequency cutoff of the form
 \begin{equation}
\label{eq:matsubaracutoff}
\hat{G}_{d,0}^\Lambda(\omega) = \Theta\left(|\omega|-\Lambda\right)
\hat{G}_{d,0}(\omega)  
\end{equation} 
with $\Lambda_0 \to \infty$.\cite{hedden04}
Evaluating $\hat{\mathcal S}$ 
by means of the Morris lemma\cite{hedden04,morris:94} results in
\begin{equation}\label{eq:single_scale}
\hat{\mathcal S}^{\Lambda}(\omega) \rightarrow \delta(|\omega|-\Lambda)
\frac{1} {\left[\hat{G}_{d,0}(\omega)\right]^{-1}-\hat\Sigma^{\Lambda}}\;\;.
\end{equation}
A straightforward calculation permits us thus to rewrite Eq.\ (\ref{eq:pot})
as
\begin{widetext}  
\begin{equation}
\label{ap}
\frac{d}{d\Lambda} \Sigma^{\mp\mp,\Lambda}=\pm \frac{iU}{2\pi}\sum_{\omega=\pm \Lambda}
\frac{  \frac{G^{\mp\mp}_{d,0}(\omega)}
{\Delta(\omega)}  - \Sigma^{\pm\pm,\Lambda}}
{ \left(\frac{G^{++}_{d,0}(\omega)}{\Delta(\omega)}  - \Sigma^{--,\Lambda}\right) 
\left(\frac{G^{--}_{d,0}(\omega)}{\Delta(\omega)}  - \Sigma^{++,\Lambda}\right)-
\left(\frac{G^{-+}_{d,0}(\omega)G^{+-}_{d,0}(\omega)}
{\Delta(\omega)^2}\right) }\;,
\end{equation}
\end{widetext}
where 
\begin{eqnarray*}
\Delta(\omega)&=&G^{--}_{d,0}(\omega)
G^{++}_{d,0}(\omega)-G^{-+}_{d,0}(\omega)
G^{+-}_{d,0}(\omega)\\
&=& -\frac{1}{(\omega-V_G)^2+\Gamma^2}\;\;.
\end{eqnarray*}
Finally, the initial condition for the self-energy is $\lim\limits_{\Lambda_0\to\infty}\hat{\Sigma}^{\Lambda_0}=0$.\cite{hedden04}

\subsubsection{Equilibrium}
 
\begin{figure}[tb]
  \begin{center}
    \includegraphics[width=0.4\textwidth,clip]{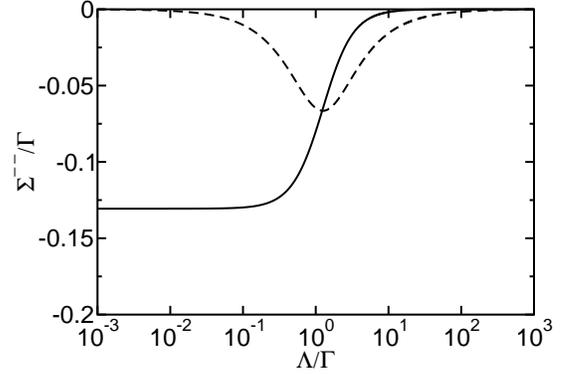}
  \end{center}
  \caption{Flow of $\Sigma^{--,\Lambda}/\Gamma$ with
    $\Lambda/\Gamma$ for $U/\Gamma=1$, $V_G/\Gamma=0.5$, and $V_B=0$. The full
    curve shows the real part, the dashed the imaginary part of
    $\Sigma^{--,\Lambda}$.}
  \label{fig:feq_ok}
\end{figure}

We now focus on $T=0$. 
In a first step we discuss the equilibrium situation, that 
is $V_B=\mu_L-\mu_R \to 0$. Then we obtain the decoupled system
\begin{equation}\label{eq:ap1}
\frac{d}{d\Lambda}\Sigma^{\mp\mp,\Lambda}= i\frac{U}{\pi}
\frac{V_G \pm \Sigma^{\mp\mp,\Lambda} }
{ \left[ (\Lambda \pm i\Gamma)^2-(V_G \pm \Sigma^{\mp\mp,\Lambda})^2  
\right]},
\end{equation}
which can be solved analytically. We first note that with
$\Sigma^{++,\Lambda}=-\left[\Sigma^{--,\Lambda}\right]^*$ both equations are
equivalent. For $\Sigma^{--,\Lambda}$ we obtain with the definition
$\sigma^\Lambda=V_G+\Sigma^{--,\Lambda}$ the solution
\begin{equation}\label{eq:analytical_solution}
\frac{i\sigma^\Lambda J_1(\frac{\pi\sigma^\Lambda}{U})-(\Lambda+i\Gamma)
  J_0(\frac{\pi\sigma^\Lambda}{U})}{i\sigma^\Lambda Y_1(\frac{\pi\sigma^\Lambda}{U})-(\Lambda+i\Gamma) Y_0(\frac{\pi\sigma^\Lambda
     }{U})}=\frac{J_0(\frac{\pi V_G}{U})}{Y_0(\frac{\pi V_G}{U})}\;,
\end{equation}
where $J_n$ and $Y_n$  are the Bessel functions of first and second kind.
The desired solution of the cutoff free problem is obtained by setting $\Lambda=0$, i.e.\ 
\begin{equation}
  \label{eq:analytic0}
\frac{\sigma J_1(\frac{\pi\sigma}{U})-\Gamma
  J_0(\frac{\pi\sigma}{U})}{\sigma Y_1(\frac{\pi\sigma}{U})-\Gamma Y_0(\frac{\pi\sigma
     }{U})}=\frac{J_0(\frac{\pi V_G}{U})}{Y_0(\frac{\pi V_G}{U})}\;.
\end{equation}
which is precisely the result Eq.~(4) obtained by \textcite{meden06:1}
It is, however, important to note that in this work, based on the
imaginary-time formulation of the fRG, the differential equation has a
different structure. It is real and has a positive definite
denominator. Thus, while the solutions at $\Lambda=0$ are identical
for the imaginary-time and real-time formulations, the flow towards
$\Lambda=0$ will show differences. As we will see next, the complex nature
of the differential equation (\ref{eq:ap1}) can lead to 
problems connected to its
analytical structure when attempting a numerical solution.
For small $U/\Gamma$ no particular problems arise.
As an example the result for the flow of
$\Sigma^{--,\Lambda}$ as function of $\Lambda$  for $U/\Gamma=1$ and
$V_G/\Gamma=0.5$ obtained with a standard Runge-Kutta solver is shown in
Fig.~\ref{fig:feq_ok}. Consistent with the analytical solution
Eq.\ (\ref{eq:analytic0}), the imaginary part (dashed line) goes to zero
as $\Lambda\to0$, while the real part (solid line) 
rapidly approaches the value
given by formula (\ref{eq:analytic0}).

\begin{figure}[tb]
  \begin{center}
    \includegraphics[width=0.4\textwidth,clip]{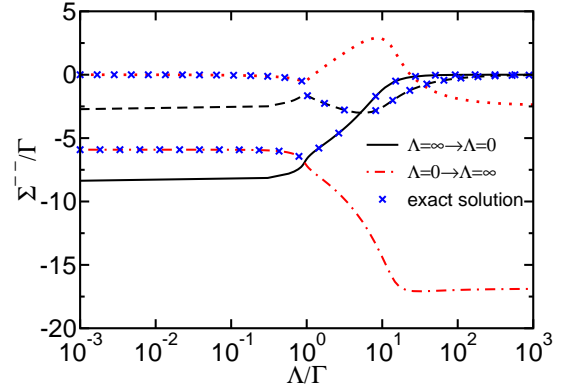}
  \end{center}
  \caption{(color online) Flow of $\Sigma^{--,\Lambda}/\Gamma$ with
    $\Lambda/\Gamma$ for $U/\Gamma=15$, $V_G/\Gamma=6$, and $V_B=0$. 
   The full and dashed
    curves show real and imaginary part obtained from the integration
    $\Lambda=\infty\to\Lambda=0$, the dashed-dotted and dotted curves
    real and imaginary part obtained from an integration
    $\Lambda=0\to\Lambda=\infty$, using the solution from
    (\ref{eq:analytic0}) as initial value for $\Sigma^{--,\Lambda}$. The
    crosses denote the analytical solution (\ref{eq:analytical_solution}).}
  \label{fig:feq_not_ok}
\end{figure}

\begin{figure}[tb]
  \begin{center}
    \includegraphics[width=0.4\textwidth,clip]{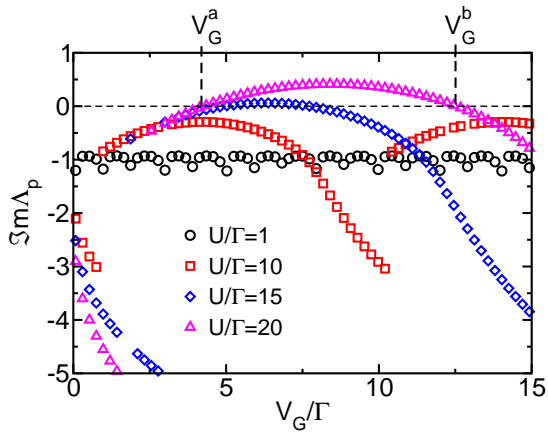}
  \end{center}
  \caption{(color online) Imaginary part of $\Lambda_p$ determined from
    (\ref{eq:analytical_solution}) and
    $V_G+\Sigma^{--,\Lambda_p}=\Lambda_p+i\Gamma$ for different values
    of $U$ and $V_B=0$ as function of $V_G$. For $U/\Gamma=15$ and $20$
    there exist an interval $[V_G^a,V_G^b]$ where $\Im m\Lambda_p>0$,
    while for small $U$ or $V_G\not\in[V_G^a,V_G^b]$ we always have $\Im m\Lambda_p<0$.}
  \label{fig:solution}
\end{figure}

However, for larger values of $U/\Gamma$ the numerical solution
becomes unstable in a certain regime of $V_G$. 
A typical result in such a situation is shown in
Fig.~\ref{fig:feq_not_ok}. The different curves were obtained as
follows: The full and dashed ones from the numerical solution starting with
$\Sigma^{--,\Lambda_0}=0$ at $\Lambda_0\to\infty$, the dash-dotted and dotted by integrating the
differential equation (\ref{eq:ap1}) backwards from $\Lambda=0$ with the correct
solution for $\Lambda=0$ as given by formula
(\ref{eq:analytic0}) as initial value. The crosses finally are the results from the analytical solution
Eq.\ (\ref{eq:analytical_solution}). Evidently, there exists a crossing of
different branches of solutions to the differential equation for
$\Lambda/\Gamma\approx1$ and the numerical solution with starting point
$\Lambda=\infty$ picks the wrong one as $\Lambda\to0$.
The reason for this behavior is that for large $U$ there exists a
certain $V_G^a$ such that
$V_G^a+\Sigma^{--,\Lambda_p}=\Lambda_p+i\Gamma$ with real $\Lambda_p$,
resulting in a pole in the differential equation (\ref{eq:ap1}).
For $V_G\ne V_G^a$ this pole does not appear for real $\Lambda_p$, but
as shown in Fig.~\ref{fig:solution} $\Im m\Lambda_p$ changes sign at $V_G^a$, which in turn induces
a sign change on the right hand side of the differential equation,
leading to the behavior observed in Fig.~\ref{fig:feq_not_ok}. 
There also exists a second critical value $V_G^b$ such that for
$V_G>V_G^b$ we find $\Im m\Lambda_p<0$ and the instability has vanished again.

Obviously, this instability limits the applicability of the present
approximation to sufficiently small values of $U$. This is different from the
imaginary-time approach by Andergassen {\em et al.},\cite{meden06:1} where this
simple approximation leads to qualitative correct results even for values
of $U$ significantly larger than $\Gamma$.

\subsubsection{Nonequilibrium}

\begin{figure}[bt]
  \begin{center}
    \includegraphics[width=0.4\textwidth,clip]{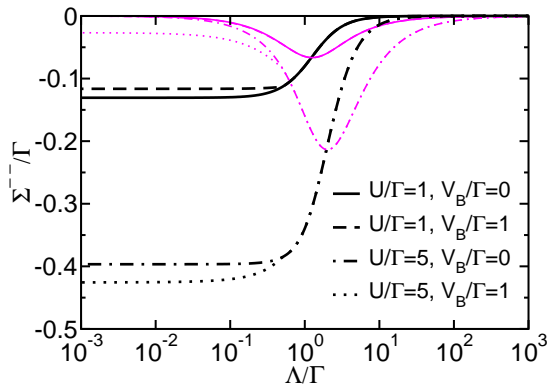}
  \end{center}
  \caption{(color online) Flow of $\Sigma^{--,\Lambda}/\Gamma$ with
    $\Lambda$ for $U/\Gamma=1$ and $5$ for $V_G/\Gamma=0.5$
 at $V_B/\Gamma=0$ and $1$ (thick curves: real part; 
thin curves: imaginary part). The curve for $\Im m
\Sigma^{--,\Lambda}/\Gamma$ at $U/\Gamma=1$ and
$V_B/\Gamma=1$ (thin dashed line) lies on top of the corresponding
zero bias curve and is thus not visible.}
  \label{fig:noneq_simple}
\end{figure}

We now turn to the case of finite bias voltage $V_B$. As a typical
example, the flow of
$\Sigma^{--,\Lambda}$ for $U/\Gamma=1$ (full and dashed curves) and $5$ (dashed-dotted
and dotted curves) for $V_G/\Gamma=0.5$ at $V_B/\Gamma=0$
(equilibrium) and $V_B/\Gamma=1$ is shown in
Fig.~\ref{fig:noneq_simple}.
Since the results for $\Sigma^{++,\Lambda}$ are related to those for $\Sigma^{--,\Lambda}$
by $\Sigma^{++,\Lambda}=-\left[\Sigma^{--,\Lambda}\right]^*$ we do not show them here.
The $V_B$ dependence of the curves for 
$\Re e\left[\Sigma^{--,\Lambda}\right]$ (thick lines) looks 
sensible. For $V_B \neq 0$ an imaginary part of order $U^2$ is generated
in the flow which does not vanish for $\Lambda \to 0$
(see the thin dotted  line). Causality requires that the relation
\begin{equation}
  \label{eq:causality}
  \Sigma^{--}(\omega)+\Sigma^{++}(\omega)
=-\left[\Sigma^{-+}(\omega)+\Sigma^{+-}(\omega)\right]
\end{equation}
must hold for the exact solution. Because of
$\Sigma^{-+}(\omega)=\Sigma^{+-}(\omega)=0$, the finite imaginary part
of $\Sigma^{\alpha\alpha}$ leads to a breaking of the condition
(\ref{eq:causality}) to order $U^2$ at the end of the fRG flow.  
This is consistent with the fact 
that by neglecting the flow of the vertex terms of 
order $U^2$ are only partially kept in the present fRG truncation
scheme. To avoid any confusion we emphasize that our RG method is
different from any low-order perturbation theory. 
The weak breaking of causality can also be understood as a 
consequence of our approximation leading to a 
{\em complex}, energy-independent self-energy: The off-diagonal 
components, being related to the distribution
functions for electrons and holes, respectively, in general
have different support on the energy axis. The energy independence
makes it impossible to respect this structure here.

For our further discussion the order $U^2$ 
violation of Eq.\ (\ref{eq:causality}) means
that we may not rely on relations like Eqs.\ (\ref{eq:Gret})-(\ref{eq:Gkeldysh})  but have to
work with $G^{\alpha\beta}$, thus the somewhat unusual formula
(\ref{eq:current}). A naive
application of $\Sigma^R=\Sigma^{--}-\Sigma^{-+}$ and use of
$G^{+-}_d-G^{-+}_d=2i \Im m G^R_d$ would have led to unphysical results.
That working with $G^{+-}_d-G^{-+}_d$ is still
sensible can be seen from a straightforward evaluation leading to
\begin{eqnarray}\label{eq:gpmmgmp_simple}
&& G^{+-}_d(\omega)-G^{-+}_d(\omega) = \\*
&&
-2i\frac{\Gamma}{\left|\omega-V_G+i\Gamma\left[1-F(\omega)\right]-\Sigma^{--}\right|^2+\Gamma^2F(\omega)F(-\omega)}\nonumber
\\[5mm]
&& F(\omega)= f_L(\omega)+f_R(\omega)\nonumber
\end{eqnarray}
which is purely imaginary with a definite sign. 
Inserting the expression (\ref{eq:gpmmgmp_simple}) into the formula
(\ref{eq:current}), one can calculate the current and thus 
the conductance. Since we are at
$T=0$, an explicit expression for the current of the cutoff free 
problem (at $\Lambda=0$) can be obtained by noting that
with $\mu_L=V_B/2$, $\mu_R=-V_B/2$ one has 
$f_L(\omega)-f_R(\omega)=\Theta(V_B/2-|\omega|)$ and 
$F(\pm\omega)=1$ for 
$\omega\in[-V_B/2,V_B/2]$, which leads to
\begin{eqnarray}
J&=&\frac{\Gamma^2}{\pi}\int\limits_{-V_B/2}^{V_B/2}d\omega
\frac{1}{|\omega-V_G-\Sigma^{--}|^2+\Gamma^2}\nonumber\\
\label{eq:current_simple}
&=&\frac{\Gamma}{\pi}\frac{\Gamma}{\Gamma^\ast}\sum\limits_{s=\pm1}s\,\arctan\left(\frac{V_G^*+s\frac{V_B}{2}}{\Gamma^\ast}\right)
\end{eqnarray}
with the abbreviations 
\begin{eqnarray}
\label{eq:abbriv_VGstar}
V_G^* &=&V_G+\Re e\Sigma^{--} \; ,\\[3mm]
\label{eq:abbriv_alpha}
\Gamma^\ast &=& \sqrt{\Gamma^2+(\Im m\Sigma^{--})^2}\;.
\end{eqnarray}
Equation (\ref{eq:current_simple}) for the current is equivalent to
the noninteracting expression but with renormalized
parameters $V_G^\ast$ and $\Gamma^\ast$, which depend on the
interaction as well as the bias and gate voltage. 

\begin{figure}[tb]
  \begin{center}
    \includegraphics[width=0.4\textwidth,clip]{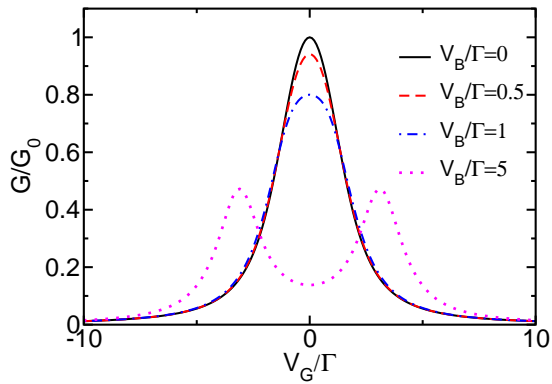}
  \end{center}
  \caption{(color online) Conductance normalized to $G_0=2e^2/h$ as function of $V_G$ for
    $U/\Gamma=2$ and several values of the bias voltage $V_B$.}
  \label{fig:conductance_simple}
\end{figure}

An example for the differential conductance as function of $V_G$  
obtained from Eq.\ (\ref{eq:current_simple})
for $U/\Gamma=2$ and several values of
$V_B$ is shown in Fig.~\ref{fig:conductance_simple}, where
$G_0=2e^2/h$ (after reintroducing $e$ and $\hbar$). 
Increasing $V_B$ leads, as expected, first to a decrease of the
conductance close to $V_G=0$ and later to a splitting of order $V_B$. 
Since we will discuss a more refined scheme including parts of the flow of
the two-particle vertex next, we do not intend to dwell too much on the results of
this simplest approximation. We note in passing that for
$V_G=0$ due to particle-hole symmetry we obtain from the differential
equation (\ref{ap}) that $\Sigma^{--}=0$ independent of $U$. Consequently
the current $J$ calculated via Eq.\ (\ref{eq:current_simple}) and
the conductance are independent of $U$,
too, and given by the corresponding expressions for the
noninteracting system. As we will see in the next section, this
deficiency will be cured by the approximate inclusion of the vertex
flow. In the present approximation the current conservation $\Delta J=0$
holds for all parameters as 
$\Sigma^{-+}=\Sigma^{+-}=0$ [cf.\ Eq.~(\ref{eq:DeltaJ2})].

\subsection{Flowing vertex}

A more refined approximation is obtained when we insert the flowing two-particle
vertex $\gamma_2^\Lambda$ as given by expression (\ref{eqq2}) in the calculation of the
self-energy Eq.\ (\ref{eq:gamma1f2}). By this we introduce an
energy-dependence of the self-energy.\cite{hedden04} However, because the size of
the resulting system of differential equations becomes extremely large
if the full frequency dependence is kept (for a discussion on this in
equilibrium see Ref.\ \onlinecite{hedden04}),
we only keep the flow of the frequency independent part of the vertex, 
an additional approximation which has successfully been used in 
equilibrium.\cite{meden06:2} As a consequence we again end up with a frequency
independent $\hat\Sigma^\Lambda$. 
The resulting expression for the self-energy (see Appendix B) is 
\begin{equation}
  \label{eq:sigme_full}
  \frac{d}{d\Lambda}\Sigma^{\alpha\beta,\Lambda}
=
-\frac{1}{2\pi}\sum\limits_{\omega= \pm \Lambda}G_d^{\gamma\delta,\Lambda}(\omega)
\left(2U^{\alpha\delta\beta\gamma,\Lambda}-U^{\delta\alpha\beta\gamma,\Lambda}\right)\;,
\end{equation}
where 
$U^{\alpha\delta\beta\gamma,\Lambda}$ is the flowing interaction given
by the expression (\ref{eqq3}).
As has been observed by Karrasch {\em et al.},\cite{meden06:2} this approximation
leads to a surprisingly accurate description of the transport properties in
equilibrium. In particular it is superior to the
lowest order approximation including  only the bare vertex. 

\begin{figure}[tb]
  \begin{center}
    \includegraphics[width=0.4\textwidth,clip]{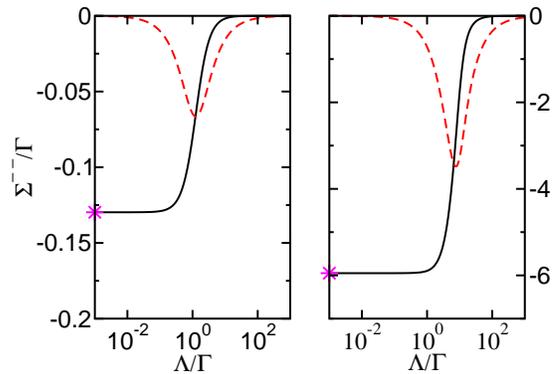}
  \end{center}
  \caption{(color online) Flow of $\Sigma^{--,\Lambda}/\Gamma$ with
    $\Lambda$ for $U/\Gamma=1$ (left panel) and $U/\Gamma=15$ (right panel)
    at $V_G=U/2$ and $V_B=0$. The full
    curves show the real part, the dashed the imaginary part of
    $\Sigma^{--,\Lambda}$. The stars at the vertical axis denote the values as
  obtained from the imaginary-time fRG.\cite{meden06:2}}
  \label{fig:full_eq}
\end{figure}

\subsubsection{Equilibrium} 

We again begin with the discussion of the
solution to Eq.\ (\ref{eq:sigme_full}) in equilibrium. 
Results for the flow of $\Sigma^{--,\Lambda}$ are presented in Fig.~\ref{fig:full_eq}
for $U/\Gamma=1$ (left panel) and $U/\Gamma=15$ (right panel)
for $V_G=U/2$. Since $\Sigma^{++,\Lambda}=-\left[\Sigma^{--,\Lambda}\right]^*$ 
only one component is
shown. The stars in Fig.~\ref{fig:full_eq} denote the solutions of the imaginary-time 
equations taken from Ref.~\onlinecite{meden06:2}. Note that for $U/\Gamma=15$
and $V_G/\Gamma>6$ the simple
approximation Eq.\ (\ref{ap}) showed an instability, while with the flowing vertex
the system is stable even for these large values of $U$ and reproduces the
correct equilibrium solutions for $\Lambda\to0$.\cite{meden06:2} The
reason for this is that
the flow of the vertex reduces the resulting effective interaction
below the critical value in the instability region.\cite{meden06:2} 

\begin{figure}[tb]
  \begin{center}
    \includegraphics[width=0.4\textwidth,clip]{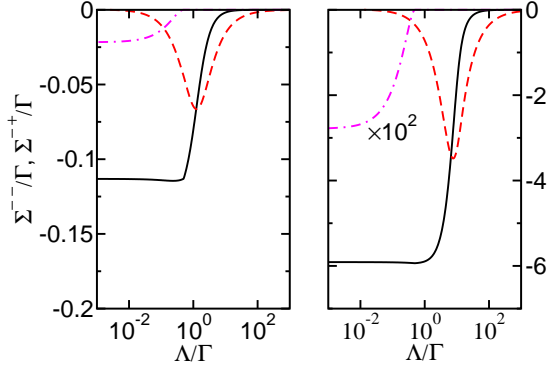}
  \end{center}
  \caption{(color online) Flow of $\Sigma^{--,\Lambda}/\Gamma$ and
    $\Sigma^{-+,\Lambda}/\Gamma$ with $\Lambda$ 
    for $U/\Gamma=1$ (left panel) and $U/\Gamma=15$ (right panel),
    $V_G/\Gamma=U/2$ and $V_B/\Gamma=1$. The full
    curves show the real part, the dashed the imaginary part of
    $\Sigma^{--,\Lambda}$, the dot-dashed the imaginary part of
    $\Sigma^{-+,\Lambda}$. The real part for the latter is zero.}
  \label{fig:full_noneq}
\end{figure}

\subsubsection{Nonequilibrium}

For the same parameters as in Fig.~\ref{fig:full_eq} we present the
resulting flow with finite bias $V_B/\Gamma=1$ in
Fig.~\ref{fig:full_noneq}.
In addition to the curves for real (solid lines) 
and imaginary part (dashed lines) of $\Sigma^{--}$
a third curve is displayed, the imaginary part of $\Sigma^{-+}$
(dashed-dotted lines), which
now is generated during the flow. Note that $\Re e\Sigma^{-+,\Lambda}=0$ and
$\Sigma^{+-,\Lambda}=\left[\Sigma^{-+,\Lambda}\right]^*$. Furthermore, we always find
$\Im m\Sigma^{-+,\Lambda}<0$. For $U/\Gamma=15$ (right panel in
  Fig.~\ref{fig:full_noneq}) we have rescaled $\Im m\Sigma^{-+,\Lambda}$ by a
  factor $10^{2}$ to make it visible on the scale of $\Sigma^{--,\Lambda}$.
Since $\Sigma^{\alpha\beta,\Lambda}$ is a complex energy independent quantity 
Eq.\ (\ref{eq:causality}) is again not fulfilled. 
We note that the error is still of order $U^2$, but for fixed
$V_G$ and $V_B$ it is significantly smaller than in the simplest truncation
scheme discussed above. 

The energy-independence of the self-energy 
allows to derive an analytical expression for the current
at $T=0$ similar to Eq.~(\ref{eq:current_simple}), which due to the
appearance of $\Sigma^{-+}$ now becomes
\begin{equation}\label{eq:current_full}
J=\frac{\Gamma}{\pi}\frac{\tilde\Gamma}{\Gamma^\ast}\sum\limits_{s=\pm1}
s\,\arctan\left(\frac{V_G^*+s\frac{V_B}{2}}{\Gamma^\ast}\right)
\end{equation}
with $V_G^*$ as in Eq.\ (\ref{eq:abbriv_VGstar}) and
\begin{eqnarray*}
  \tilde\Gamma & = & \Gamma-\Im m\Sigma^{-+}>\Gamma\\[3mm]
\Gamma^\ast & = & \sqrt{\tilde\Gamma^2+(\Im m\Sigma^{--})^2}\;;
\end{eqnarray*}
where $\Sigma^{\alpha\beta}$ is taken at $\Lambda=0$.
Thus, the only change to the expression (\ref{eq:current_simple}) is a formal
replacement $\Gamma\to\tilde\Gamma$ in $ J/\Gamma$. Equation
(\ref{eq:current_full}) 
is of the same structure as for the noninteracting case with $V_G$ and $\Gamma$
replaced by renormalized parameters. However, the two self-energy
contributions $\Sigma^{--}$ and $\Sigma^{-+}$ enter
distinctively different in the expression for the current. While $\Im m\Sigma^{--}$
solely plays the role of an additional life-time broadening, $\Im
m\Sigma^{-+}$ directly modifies the tunneling rate
both in the prefactor of $J$ {\em and} in the expression for the life-time broadening.

A problem occurs when using the results of the present 
approximation in Eq.~(\ref{eq:DeltaJ2}), leading to
\begin{widetext}
\begin{equation}
  \label{eq:DeltaJAppr}
  \Delta J =
2\frac{\Gamma}{\pi}\int\limits_{-\infty}^\infty d\omega
\frac{\Im m\Sigma^{-+}\left[1-F(\omega)\right]}{
\left|\omega-V_G+i\Gamma\left[1-F(\omega)\right]-\Sigma^{--}\right|^2+\left[\Gamma
  F(-\omega)+\Im m\Sigma^{+-}\right]\left[\Gamma
  F(\omega)-\Im m\Sigma^{-+}\right]}\;\;.
\end{equation}
\end{widetext}
The requirement $\Delta J=0$ is only fulfilled for $V_G=0$, because
then $\Sigma^{--}=0$ and the integrand is asymmetric with respect to
$\omega$. Thus our approximation of an
energy-independent flowing vertex violates current conservation for
$V_G\ne0$ in nonequilibrium. 
We verified that $\Delta J \sim U^2$ which is
consistent with the fact that not all terms of order $U^2$ are kept in
our truncated fRG procedure.  
How does $\Delta J$ 
behave in the limit $V_B\to0$? To see this
we  note that, because
$\Im m\Sigma^{-+}$ does not depend on the sign of $V_B$ and
furthermore goes to zero as $V_B\to0$, $\Im
m\Sigma^{-+}\stackrel{V_B\to0}{\sim}V_B^2$. Consequently, $\Delta
J\stackrel{V_B\to0}{\sim}V_B^2$ and hence the violation of current
conservation vanishes in the linear response regime $V_B\to0$.

\begin{figure}[tb]
  \begin{center}
    \includegraphics[width=0.4\textwidth,clip]{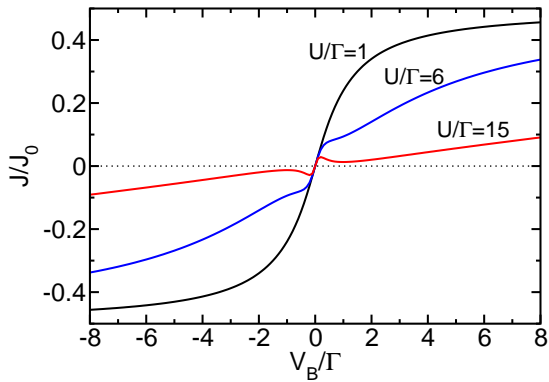}
  \end{center}
  \caption{(color online) Current normalized to
    $J_0=G_0\frac{\Gamma}{e}$ (after reintroducing $e$ and $\hbar$) 
as function of $V_B$ for $U/\Gamma=1$,
$6$ and $15$ and $V_G=0$. For $U/\Gamma=15$ we find a region of negative differential
conductance in the region $|V_B/\Gamma|\approx0.5$ (c.f.\ Fig.\ \ref{fig:G_Vs_mu}).}
  \label{fig:current_full}
\end{figure}

\begin{figure}[tb]
  \begin{center}
    \includegraphics[width=0.4\textwidth,clip]{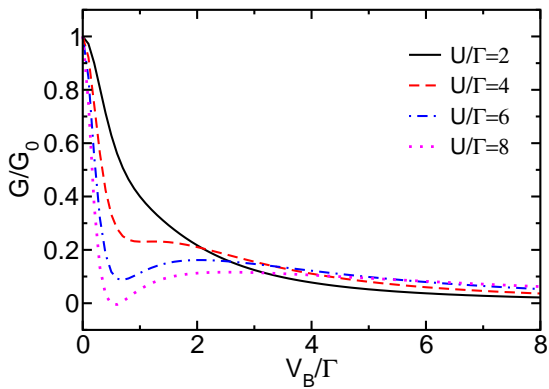}
  \end{center}
  \caption{(color online) Differential conductance $G$ as function of $V_B$ for $V_G=0$ and various values of $U$. For $U/\Gamma>5$ a distinct minimum around $V_B/\Gamma\approx0.5$ appears.}
  \label{fig:G_Vs_mu}
\end{figure}

In Fig.~\ref{fig:current_full}
we show the current at $V_G=0$ as function of $V_B$ for $U/\Gamma=1$, $6$ and
$15$. With increasing $U$ the current for intermediate $V_B$ is strongly suppressed.
In addition there occurs a structure at low $V_B$,
which turns into a region of negative differential conductance with increasing $U$.
The appearance of such a shoulder in the current
was observed in other
calculations as well.\cite{Hershfield:1991,hershfield:92,fujii:03} 
However, whether the
negative differential conductance we find for still larger values of
$U$ (c.f.\ Fig.\ \ref{fig:G_Vs_mu}) is a true feature of the
model or rather an artifact of the approximations used is presently
not clear and should be clarified in further investigations.
However, negative differential conductance has also been observed
in a slave-boson treatment of the model.\cite{takahashi:2006}

Keeping $V_G=0$ fixed, we can calculate the conductance $G=dJ/dV_B$ as function
of $V_B$ for different values of $U$. The results are collected in Fig.~\ref{fig:G_Vs_mu}.
In contrast to the simple approximation without flow of the vertex, the 
conductance is now strongly dependent on $U$, except for $V_B=0$, where due
to the unitary limit at $T=0$ we always find $G=G_0$. As already anticipated from
the current in Fig.~\ref{fig:current_full}, a minimum in $G$ starts to form
around $V_B/\Gamma\approx0.5$ for $U/\Gamma>5$, which is accompanied by
a peak at $V_B/\Gamma\approx2$. A similar behavior in the conductance
was observed in a perturbative treatment,\cite{fujii:03}  which in contrast to
our current approximation involves the full energy-dependence in the
self-energy.
This at least qualitative agreement -- we of course cannot resolve
structures like the Hubbard bands with an energy independent
self-energy -- again supports our claim that despite the
violation of the relation (\ref{eq:causality}) we can obtain reasonable
results from $G^{\alpha\beta}$.

\begin{figure}[tb]
  \begin{center}
    \includegraphics[width=0.4\textwidth,clip]{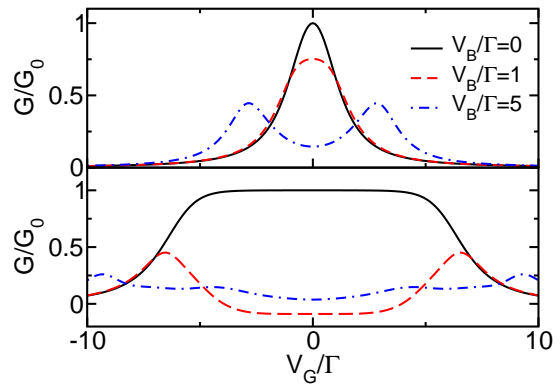}
  \end{center}
  \caption{(color online) Differential conductance $G$ as function of $V_G$ for different values of $V_B$ and $U/\Gamma=1$ (upper panel) and $U/\Gamma=15$ (lower panel). Note the extended plateau at $V_B=0$ for $U/\Gamma=15$, which is a
manifestation of the pinning of spectral weight at the Fermi level.}
  \label{fig:G_Vs_VG}
\end{figure}

We finally discuss the variation of the conductance
with $V_G$ for fixed $U$ and $V_B$. We again emphasize, that for
$V_G\ne0$, $\Delta J=0$ only holds to leading order in $U$.
In Fig.~\ref{fig:G_Vs_VG} we present the curves for two different values of $U$, 
namely $U/\Gamma=1$ (upper panel in Fig.~\ref{fig:G_Vs_VG}) and $U/\Gamma=15$
(lower panel in Fig.~\ref{fig:G_Vs_VG}). In the former case, the variation
of $G$ with $V_B$ is rather smooth,
as is to be expected from the current
in Fig.~\ref{fig:current_full}. For large $U$, we observe an extended plateau
at zero bias, which is a manifestation of the fact that in the strong coupling
regime a pinning of spectral weight at the Fermi energy occurs. 
This feature is also observed in the imaginary-time fRG as
well as in NRG calculations.\cite{meden06:2}
Increasing $V_B$ quickly leads to a similarly extended region of
negative differential conductance,
which, assuming that this result is a true feature of the model,
therefore seems to be linked to the ``Kondo'' pinning. 
We note that it is unlikely that the appearance of the negative
differential conductance
is related to the breaking of current
conservation at order $U^2$ as it also appears for $V_G=0$ where
$\Delta J=0$. For large $V_B$ multiple
structures appear in $G$, which are related to the energy scales $V_B$ and $U$.

\section{Summary and outlook}
\label{summary}
Starting from a generating functional proposed by
Kameneev\cite{kameneev} we have derived an infinite hierarchy of
differential equations for the vertex functions of an interacting
quantum mechanical many-body system in nonequilibrium (see also
Refs.~\onlinecite{jakobs:03} and \onlinecite{jakobsneu}). The major
difference to the imaginary-time formulation comes from
the use of Keldysh Green functions. The indices $+$ and $-$ 
referring to the upper and lower branch of the (real-)time contour 
lead to an additional matrix structure to the problem. Our 
formulation is sufficiently general that it allows to treat bosonic 
and fermionic models with or without explicit time dependence and 
at $T\ge0$. Since the
fRG leads to an infinite hierarchy of coupled differential
equations, one has to introduce approximations, at least a truncation
at a certain level. Typically one neglects the flow of the
three-particle vertex. As has been demonstrated in
Ref.~\onlinecite{hedden04} for the imaginary-time fRG one can solve the
remaining system of flow equations for simple models like the SIAM
numerically, thus keeping the full energy-dependence. Due to
the fact that the vertex function carries three {\em continuous}
frequency arguments in addition to the discrete quantum numbers of the
system, such a calculation can become computationally quite
expensive. 

To reduce the numerical effort, further additional approximations can
be introduced. A particularly important and successful one is obtained 
by neglecting the energy dependence of the vertex 
functions,\cite{meden06:1,meden06:2} which already leads to a
surprisingly accurate description of local and  
transport properties of interacting quantum dots in the linear
response regime. 

We applied the nonequilibrium fRG to the SIAM
with finite bias voltage in the stationary state. It turned out, that for
the simplest approximation where only the flow of the self-energy is
kept, the analytic structure of the differential equation leads to problems in the
numerical solution. In addition, this approximation leads to a violation
of the causality relation (\ref{eq:causality}) to order $U^2$. 
The first problem was resolved by including the 
two-particle vertex in the flow at least up to the largest
interaction considered here ($U/\Gamma =15$). At the 
present stage this was for computational reasons done by assuming it
to be energy-independent, yielding again an energy-independent
self-energy. Although this approximation also violates 
Eq.\ (\ref{eq:causality}) to order $U^2$ for a fixed $V_G$ and $V_B$ the
error is significantly smaller compared to the simplest scheme. 
We were able to obtain
reasonable expressions and numerical results for the current and the
conductance using the functions $G^{\alpha\beta}(\omega)$ instead of
$G^{R}(\omega)$ in the current formula. We reproduced 
nonequilibrium features of the current and differential conductance 
known from the application of other approximate methods to the SIAM. 

In the more advanced truncation scheme and for $V_G \neq 0$, 
the current conservation $\Delta J = 0$ only holds to leading order in $U$. 
This defect can be traced back to the
energy independence of the two-particle vertex, leading to finite, but
energy-independent $\Sigma^{-+}$ and $\Sigma^{+-}$. By a close
inspection of formula (\ref{eq:DeltaJ2}), however, one can see
that this deficiency can for example be cured by assuming a
coarse-grained energy dependence of the form
\begin{widetext}
\begin{eqnarray}
  \label{minmaleEnergie}
\omega<-\frac{V_B}{2}: &F(\omega)=2\;\Rightarrow& \Im
m\Sigma^{+-,\Lambda}=0\;\mbox{, $\Im m\Sigma^{-+,\Lambda}$=const.}\\[3mm]
-\frac{V_B}{2}<\omega<\frac{V_B}{2}: &F(\omega)=1\;\Rightarrow& \Im
m\Sigma^{+-,\Lambda}=-\Im m\Sigma^{-+,\Lambda}=\mbox{const.}\\[3mm]
\frac{V_B}{2}<\omega: &F(\omega)=0\;\Rightarrow& \Im
m\Sigma^{-+,\Lambda}=0\;\mbox{, $\Im m\Sigma^{+-,\Lambda}$=const.}
\end{eqnarray}
\end{widetext}
and corresponding energy dependencies for $\Sigma^{--}$ and
$\Sigma^{++}$. Here the const. might depend on $\Lambda$ but not on
$\omega$. Such an approximation will quite likely also
reestablish causality Eq.\ (\ref{eq:causality}). Work along this
line is in progress.

Including finite temperature, magnetic field etc.\ in the calculations
of transport is easily possible, as well as the extension to more
complicated ``impurity'' structures (see e.g.\
Ref.~\onlinecite{meden06:2}). The latter aspect
is typically a rather cumbersome step for other methods like
perturbation theories or in particular numerical techniques. 

Our present stage of work 
should mainly be seen as a ``proof of principle''. 
Already for the SIAM there remain a
variety of fundamental things to do; like the implementation of the
full energy-dependence in the calculations, which then should cure the
violation of charge conservation and causality; or 
implementing a reasonable
``cutoff scheme'' for time-dependent problems off equilibrium, e.g.\ to
calculate current transients. Despite its current limitations we
expect the real-time nonequilibrium formulation of the fRG to be a 
useful tool to understand nonequilibrium features of
mesoscopic systems, just like the imaginary-time version in
the calculation of equilibrium properties.

\begin{acknowledgments}
We acknowledge useful conversations with
H.~Schoeller,
S.~Jakobs,
S.~Kehrein,
J.~Kroha,
H.~Monien,
A.~Schiller,
A.~Dirks,
J.~Freericks,
K.~Ueda,
T.~Fujii,
K.~Thygesen,
and
F.B.~Anders. We thank H.~Schmidt for pointing out 
typos in an earlier version of our paper. 
This work was supported by the DFG through the collaborative research
center SFB 602. Computer support was provided through the Gesellschaft
f\"ur wissenschaftliche Datenverarbeitung in G\"ottingen and
the Norddeutsche Verbund f\"ur Hoch- und H\"ochstleistungsrechnen.
\end{acknowledgments}

\appendix

\begin{widetext}
\section{Derivation of the flow equations}

Although the following derivation is mainly equivalent to the
one presented in Ref.~\onlinecite{hedden04}, the appearance of the
factor $i$ in the real-time formulation of the generating functional
leads to some changes in signs and prefactors $i$ in the equations. 
It thus appears to be helpful for the reader to repeat the derivation.

As a first step we 
differentiate ${\mathcal W}^{c,\Lambda}$ 
with respect to $\Lambda$, which after straightforward algebra leads
to 
\begin{equation}
\label{flusswc}
\frac{d}{d \Lambda} {\mathcal W}^{c,\Lambda} = 
\zeta \mbox{ Tr}\, \left( 
\hat {\mathcal Q}^{\Lambda} \hat G^{0,\Lambda}
 \right) +i\zeta 
\mbox{Tr} \, 
\left(\hat {\mathcal Q}^{\Lambda} 
 \frac{\delta^2
   {\mathcal W}^{c,\Lambda} }{ \delta \bar \eta \delta
      \eta}  \right) + \left(
   \frac{\delta {\mathcal W}^{c,\Lambda} }{\delta \eta},
   \hat {\mathcal Q}^{\Lambda}  \frac{\delta {\mathcal
       W}^{c,\Lambda} }{\delta \bar \eta} \right) \; .
\end{equation}
Considering $\phi$ and $\bar \phi$ as the fundamental variables we
obtain from Eq.\ (\ref{eq:gammadef}) 
$$
\frac{d}{d \Lambda} \Gamma^{\Lambda}\left(\{\bar \phi \}, \{ \phi \} \right) = 
- \frac{d}{d \Lambda} {\mathcal W}^{c,\Lambda} \left(\{\bar
  \eta^{\Lambda} \}, \{ \eta^{\Lambda} \} \right) 
- \left(\bar \phi, \frac{d}{d \Lambda} \eta^{\Lambda}
  \right) - \left( \frac{d}{d \Lambda} {\bar \eta}^{\Lambda} , \phi \right) 
+  \left(\bar \phi,  \hat {\mathcal Q}^{\Lambda}  \phi\right) \; .
$$
Applying the chain rule and using Eq.\ (\ref{flusswc}) this leads to 
\begin{eqnarray*}
\frac{d}{d \Lambda} \Gamma^{\Lambda} = - \zeta \mbox{ Tr}\, \left(
 \hat {\mathcal Q}^{\Lambda} 
\hat G^{0,\Lambda}  \right) 
- i \zeta\mbox{Tr} \, \left(  \hat {\mathcal Q}^{\Lambda} 
 \frac{\delta^2
   {\mathcal W}^{c,\Lambda} }{ \delta \bar \eta^{\Lambda} \delta
      \eta^{\Lambda}}  \right) \; ,
\end{eqnarray*}
where the last term in Eq.\ (\ref{eq:gammadef}) cancels a corresponding
contribution arising in (\ref{flusswc}) thus a posterior justifying
the inclusion of this term. 

Extending the well known relation\cite{negele} between the second
functional derivatives of $\Gamma$ and ${\mathcal W}^{c}$ to
nonequilibrium we obtain the functional differential equation 
\begin{equation}
\label{gammafluss}
\frac{d}{d \Lambda} \Gamma^{\Lambda} = - \zeta 
\mbox{ Tr}\, \left( \hat {\mathcal Q}^{\Lambda}  
     \hat G^{0,\Lambda} 
 \right) - \mbox{Tr} \, \left(\hat {\mathcal Q}^{\Lambda}   
 {\mathcal V}_{\bar \phi, \phi}^{1,1}(\Gamma^{\Lambda}, 
\hat G^{0,\Lambda})  \right) \; ,
\end{equation}
where $ {\mathcal V}_{\bar \phi, \phi}^{1,1}$ stands for the upper left
block of the matrix
\begin{eqnarray}
\label{invers}
{\mathcal V}_{\bar \phi,\phi}(\Gamma^{\Lambda}, \hat G^{0,\Lambda}) = 
\left( \begin{array}{cc} 
i\frac{\delta^2 \Gamma^{\Lambda}}{ \delta \bar \phi \delta \phi}
   - \zeta \left[ \hat G^{0,\Lambda} \right]^{-1}
& i \frac{\delta^2 \Gamma^{\Lambda}}{\delta \bar \phi \delta  \bar \phi} \\
 i \zeta\frac{\delta^2 \Gamma^{\Lambda}}{\delta  \phi \delta  \phi} 
& -\left\{ i\frac{\delta^2 \Gamma^{\Lambda}}{\delta \phi \delta \bar \phi} + 
\left( \left[ \hat G^{0,\Lambda} \right]^{-1}\right)^t\right\} 
\end{array} \right)^{-1} 
\end{eqnarray}
and the upper index $t$ denotes the transposed matrix.
To obtain differential equations for the $\gamma_m^{\Lambda}$ which
include self-energy corrections we express $ {\mathcal V}_{\bar \phi,
  \phi}$  in terms of
$\hat G^{\Lambda}$ instead of $\hat G^{0,\Lambda}$.
This is achieved by defining 
\begin{eqnarray*}
{\mathcal U}_{\bar \phi,   \phi} = i\frac{\delta^2 \Gamma^{\Lambda}}{\delta \bar
  \phi \delta \phi} - \gamma_1^{\Lambda}  
\end{eqnarray*}
and using
\begin{eqnarray}
\label{GG}
\hat G^{\Lambda} = \left[ \left[ \hat G^{0,\Lambda}
  \right]^{-1} -  \zeta \gamma_1^{\Lambda}  \right]^{-1} \;,
\end{eqnarray}
which leads to
\begin{eqnarray}
\label{gammaflussalt}
\frac{d}{d \Lambda} \Gamma^{\Lambda} = - \zeta \mbox{ Tr}\, \left(
  \hat {\mathcal Q}^{\Lambda}  \hat G^{0,\Lambda} 
 \right) + \zeta \mbox{Tr} \, \left[ \hat G^{\Lambda} 
   \hat {\mathcal Q}^{\Lambda} 
 \tilde{\mathcal V}_{\bar \phi, \phi}^{1,1}(\Gamma^{\Lambda},\hat
 G^{\Lambda} )  
\right]
\; , 
\end{eqnarray} 
with 
\begin{eqnarray}
\label{Vtildedef}
\tilde {\mathcal V}_{\bar \phi,\phi}
\left(\Gamma^{\Lambda}, \hat G^{\Lambda}  \right)  = \left[ {\bf 1} - 
\left( \begin{array}{cc} 
\zeta \hat G^{\Lambda} 
&  
0\\
0
&  
\left[ \hat G^{\Lambda} \right]^t 
\end{array} \right)
\left( \begin{array}{cc} 
{\mathcal U}_{\bar \phi,   \phi}
&  i\frac{\delta^2 \Gamma^{\Lambda}}{\delta \bar \phi \delta  \bar \phi} \\
(i\zeta)\frac{\delta^2 \Gamma^{\Lambda}}{\delta  \phi \delta  \phi} 
&  \zeta  {\mathcal U}^{t}_{\bar \phi,   \phi}
\end{array} \right) \right]^{-1} \; .
\end{eqnarray} 
It is important to note that ${\mathcal U}_{\bar \phi,
  \phi}$ as well as $ \frac{\delta^2
  \Gamma^{\Lambda}}{\delta \bar \phi \delta \bar \phi}$ and $\frac{\delta^2
  \Gamma^{\Lambda}}{\delta  \phi \delta  \phi} $ are at least
quadratic in the external sources. 
The initial condition for the exact functional differential equation 
(\ref{gammaflussalt}) can either be obtained by lengthy but 
straightforward algebra, which we are not going to present here or 
by the following simple 
argument: at $\Lambda= \Lambda_0$, $\hat G^{0,\Lambda_0}=0$ (no
degrees of freedom are ``turned on'') and in a perturbative expansion
of the $\gamma_m^{\Lambda_0}$ the only term which does not vanish is the
bare two-particle vertex. We thus find
 \begin{eqnarray}
\label{anfanggamma}
\Gamma^{\Lambda_0} \left( \{ \bar \phi \} , \left\{  \phi
  \right\} \right) = S_{\rm int}  \left(\{\bar
   \phi \}, \{ \phi \} \right) \; .
\end{eqnarray} 
Expanding $\tilde{\mathcal V}_{\bar \phi,\phi}$ Eq.\ (\ref{Vtildedef}) in a geometric series 
\begin{eqnarray}
\label{eq:Vtildeexp}
  \tilde{\mathcal V}_{\bar \phi,\phi} & = & {\mathbf 1} +\zeta \hat G^\Lambda{\mathcal
    U}_{\bar{\phi},\phi}+
\hat G^\Lambda {\mathcal U}_{\bar{\phi},\phi} \hat G^\Lambda{\mathcal
    U}_{\bar{\phi},\phi}
+\zeta^3i^2 \hat G^\Lambda
\frac{\delta^2\Gamma}{\delta\bar\phi\delta\bar\phi}\left[\hat G^\Lambda\right]^t
\frac{\delta^2\Gamma}{\delta\phi\delta\phi}+  \ldots
\end{eqnarray}
and
$\Gamma^{\Lambda}$ in a Taylor series with respect to the external sources
\begin{eqnarray}
\Gamma^{\Lambda}\left(\{\bar \phi \}, \{ \phi \} \right) =
\sum_{m=0}^{\infty} \frac{(i\zeta)^m}{(m !)^2}  \sum_{\xi_1', \ldots, \xi_m'} 
\sum_{\xi_1, \ldots, \xi_m } && \gamma_m^{\Lambda}\left(\xi_1',
  \ldots, \xi_m'; \xi_1, \ldots, \xi_m \right)
\label{Gumma}
\,\bar \phi_{\xi_1'} \ldots \bar
\phi_{\xi_m'} \phi_{\xi_m} \ldots  \phi_{\xi_1} \; .
\end{eqnarray}
one can derive an exact infinite hierarchy of flow equations 
for the $\gamma_m^{\Lambda}$. 
For example, the flow equation for the single-particle vertex
$\gamma_1$ (the self-energy) can be obtained by taking the expansion
(\ref{eq:Vtildeexp}) up to first order in ${\mathcal U}_{\bar \phi,\phi}$, inserting it
into (\ref{gammaflussalt}), replacing $\Gamma^\Lambda$ on both sides by
the expansion (\ref{Gumma}) and comparing the terms quadratic in the
external fields. This procedure leads to the expression (\ref{eq:gamma1fl}).

\section{Flow equation for the two-particle vertex\label{app:B}}

To obtain the flow equation for $\hat\Sigma^\Lambda$ we have
expanded Eq.\ (\ref{Vtildedef}) in a geometric series 
up to first order and $\Gamma^{\Lambda}$ (\ref{Gumma}) 
in the external sources up to $m=2$.
To find the flow equation for the vertex function,
$\gamma_2^\Lambda$, we have to proceed one step further, that is expand
 (\ref{Vtildedef}) up to the second order and $\Gamma^{\Lambda}$  
up to $m=3$. After comparison of the terms with the same
power in the fields we obtain an expression similar to the equilibrium
one (see Eq.~(15) of Ref.\ \onlinecite{meden06:2})
\begin{eqnarray}
 \frac{d}{d\Lambda} \gamma^{\alpha \beta \gamma \delta,\Lambda}(\xi_1',\xi_2';\xi_1,\xi_2)
 =\sum_{\xi_3,\xi_3'} \sum_{\xi_4,\xi_4'}\sum_{\mu,\nu,\rho,\eta}
\!\!\!  & \bigg( & \!\!\!
G^{\rho \eta,\Lambda}_{\xi_3',\xi_3} S^{\nu \mu,\Lambda}_{\xi_4,\xi_4'}\left[
\gamma^{\alpha \beta \rho \nu,\Lambda}(\xi_1',\xi_2';\xi_3,\xi_4)  
\gamma^{\eta \mu \gamma\delta,\Lambda}(\xi_3',\xi_4';\xi_1,\xi_2) \right]\nonumber\\ 
 & - &
G^{\eta \rho,\Lambda}_{\xi_3,\xi_3'} S^{\nu \mu,\Lambda}_{\xi_4,\xi_4'}\big[ 
\gamma^{\alpha \mu \gamma \eta,\Lambda}(\xi_1',\xi_4';\xi_1,\xi_3)  
\gamma^{\rho \beta \nu \delta,\Lambda}(\xi_3',\xi_2';\xi_4,\xi_2) 
\nonumber\\ 
& + &
\gamma^{\alpha\rho\gamma\nu,\Lambda}(\xi_1',\xi_3';\xi_1,\xi_4)\gamma^{\mu\beta\eta\delta,\Lambda}(\xi_4',\xi_2';\xi_3,\xi_2)\nonumber\\
& -& 
\gamma^{\beta \mu \gamma \eta,\Lambda}(\xi_2',\xi_4';\xi_1,\xi_3)\gamma^{\rho \alpha \nu
  \delta,\Lambda}(\xi_3',\xi_1';\xi_4,\xi_2)
\nonumber\\ 
\label{eqq1} & - &
\gamma^{\beta\rho\gamma\nu,\Lambda}(\xi_2',\xi_3';\xi_1,\xi_4)\gamma^{\mu\alpha\eta\delta,\Lambda}(\xi_4',\xi_1';\xi_3,\xi_2)
\big] \bigg)\;,
\end{eqnarray}
\end{widetext}
where the contribution involving $\gamma_3^\Lambda$ was replaced by its
initial value, i.e.\ $\gamma_3^\Lambda=\gamma_3^{\Lambda_0}=0$.
The main difference to equilibrium is the presence of mixed
contractions for the Keldysh indices of the matrices $\hat{G}^\Lambda$ 
and $\hat{\mathcal S}^\Lambda$ and the four-index tensors $\gamma_2^\Lambda$.

Up to this point the derivation of the flow equation for the two-particle
vertex was as general as the discussion in Sec.\ \ref{formalism}.
We now consider the application to the steady state current through
a quantum dot described by the SIAM presented in Sec.\ \ref{appli}. We
thus go over to frequency space and introduce the cutoff $\Lambda$ as in
Eq.\ (\ref{eq:matsubaracutoff}). The single-particle quantum number in
$\xi$ is given by the spin index. We can take advantage of spin and
energy conservation which implies that the lower indices of 
$\hat G$ and $\hat {\mathcal S}$ are equal, that is $\xi_3=\xi_3'$
and $\xi_4 = \xi_4'$. This does {\em not} hold for
the Keldysh indices. The integrals over  $\omega_{3'}$ and $\omega_{4'}$
as well as the sums over $\sigma_3'$ and $\sigma_4'$ can then be
performed trivially. Using the Morris lemma,\cite{morris:94}
we can rewrite the matrix product of $\hat G$ and $\hat {\mathcal S}$ as
\begin{eqnarray*}
\hat G^{\Lambda}(\omega) \hat {\mathcal S}^{\Lambda}(\omega') \rightarrow
\frac{1}{2} \delta(|\omega'|-\Lambda) \hat G_d^\Lambda(\omega)
\hat G_d^\Lambda(\omega')\;,
\end{eqnarray*}
with 
\begin{eqnarray*}
 \hat G_d^\Lambda(\omega) = \frac{1}{\left[\hat G_{d,0}(\omega)\right]^{-1} -
 \hat \Sigma^{\Lambda}(\omega)} \; .
\end{eqnarray*}
The $\delta$-function can be used to perform another of the frequency
integrals, and the remaining one can be evaluated because of energy
conservation of the two-particle vertex. Using the spin independence
of the Green function for zero magnetic field, this leads to 
\begin{widetext}
\begin{eqnarray}
 \frac{d}{d\Lambda} \gamma^{\alpha \beta \gamma
   \delta,\Lambda}_{\sigma_1',\sigma_2';\sigma_1,\sigma_2}
 = \frac{1}{4\pi}\sum_{\omega=\pm \Lambda}
 \sum_{\sigma_3,\sigma_4}\sum_{\mu,\nu,\rho,\eta} 
\bigg(&\!\!\!G_d^{\rho \eta,\Lambda}(-\omega) G_d^{\nu \mu,\Lambda}(\omega)&\!\!\!
\gamma^{\alpha \beta \rho \nu,\Lambda}_{\sigma_1',\sigma_2';\sigma_3,\sigma_4}  
\gamma^{\eta \mu \gamma\delta,\Lambda}_{\sigma_3,\sigma_4;\sigma_1,\sigma_2}
\nonumber\\
&\!\!\!-G_d^{\eta \rho,\Lambda}(\omega) G_d^{\nu \mu,\Lambda}(\omega)\Big[ &\!\!\!
\gamma^{\alpha \mu \gamma \eta,\Lambda}_{\sigma_1',\sigma_4;\sigma_1,\sigma_3}  
\gamma^{\rho \beta \nu \delta,\Lambda}_{\sigma_3,\sigma_2';\sigma_4,\sigma_2} 
+
\gamma^{\alpha \rho \gamma \nu,\Lambda}_{\sigma_1',\sigma_3;\sigma_1,\sigma_4}  
\gamma^{\mu \beta \eta \delta,\Lambda}_{\sigma_4,\sigma_2';\sigma_3,\sigma_2} \nonumber\\
\label{eqq2}
&\!\!\!&\!\!\!-
\gamma^{\beta \mu \gamma \eta,\Lambda}_{\sigma_2',\sigma_4;\sigma_1,\sigma_3}\gamma^{\rho \alpha \nu
  \delta,\Lambda}_{\sigma_3,\sigma_1';\sigma_4,\sigma_2}
-
\gamma^{\beta \rho \gamma \nu,\Lambda}_{\sigma_2',\sigma_3;\sigma_1,\sigma_4}\gamma^{\mu \alpha \eta
  \delta,\Lambda}_{\sigma_4,\sigma_1';\sigma_3,\sigma_2}
\Big] \bigg)\;\;.
\end{eqnarray}
Comparing (\ref{eqq2}) with Eq.~(20) in Ref.~\onlinecite{meden06:2},
were similar approximations were made in equilibrium, we see
that we have two more terms because of the Keldysh indices. 
In the first term appears $G_d^{\rho \eta,\Lambda}$, 
while its transpose $G_d^{\eta \rho,\Lambda}$ enters everywhere else.

Using the antisymmetry of $\gamma_2^\Lambda$ in the spin indices 
and spin conservation we can further simplify 
Eq.\ (\ref{eqq2}) by introducing the flowing interaction 
$U^{\alpha \beta \gamma \delta,\Lambda}$ defined as 
$$
\gamma^{\alpha \beta \gamma
  \delta,\Lambda}_{\sigma_1',\sigma_2';\sigma_1,\sigma_2}=\delta_{\sigma_1',\sigma_1} 
\delta_{\sigma_2',\sigma_2} 
U^{\alpha \beta \gamma \delta,\Lambda}-
\delta_{\sigma_2',\sigma_1}\delta_{\sigma_1',\sigma_2} 
U^{\beta \alpha \gamma \delta,\Lambda}\;.
$$
This leads to the flow equation
\begin{eqnarray}
\frac{d}{d\Lambda} U^{\alpha \beta \gamma \delta,\Lambda}
 = \frac{1}{4\pi}\sum_{\omega=\pm
   \Lambda} \sum_{\mu,\nu\rho,\eta} 
\biggr(&\!\!\!G_d^{\rho \eta,\Lambda}(-\omega) G_d^{\nu \mu,\Lambda}(\omega)&\!\!\!\left[
U^{\alpha \beta \rho \nu,\Lambda} 
U^{\eta \mu \gamma\delta,\Lambda}+ 
U^{\beta \alpha \rho \nu,\Lambda} U^{\mu \eta \gamma\delta,\Lambda}\right]\nonumber \\
&\!\!\!-G_d^{\eta \rho,\Lambda}(\omega) G_d^{\nu \mu,\Lambda}(\omega)&\!\!\!\big[ 
2U^{\alpha \mu \gamma \eta,\Lambda} U^{\rho \beta \nu \delta,\Lambda}- 
U^{\alpha \mu \gamma \eta,\Lambda} U^{\beta \rho \nu \delta,\Lambda}-
U^{\mu \alpha \gamma \eta,\Lambda} U^{\rho \beta \nu \delta,\Lambda}\nonumber\\
&\!\!\!&+2U^{\alpha \rho \gamma \nu,\Lambda} U^{\mu \beta \eta \delta,\Lambda}-
U^{\alpha \rho \gamma \nu,\Lambda} U^{\beta \mu \eta \delta,\Lambda}-
U^{\rho \alpha \gamma \nu,\Lambda} U^{\mu \beta \eta \delta,\Lambda}\nonumber\\
  \label{eqq3}
&\!\!\!&-
U^{\mu \beta \gamma \eta,\Lambda}U^{\alpha \rho \nu \delta,\Lambda}
-U^{\rho \beta \gamma \nu,\Lambda}U^{\alpha \mu \eta \delta,\Lambda}
\big] \bigg)\;\;.
\end{eqnarray}
\end{widetext}
From the initial value of $\gamma_2^\Lambda$ at $\Lambda = \Lambda_0
\to \infty$ given in
Eq.~(\ref{eq:vertexinit}) we can read off as the initial value for
$U^{\alpha\beta\gamma\delta,\Lambda_0}$
$$
U^{\alpha\alpha\alpha\alpha,\Lambda_0} = \alpha i U  \; ,
$$
while all other components are zero.

\end{document}